\documentclass[pdflatex,sn-mathphys-num]{sn-jnl}
\usepackage{natbib}
\theoremstyle{thmstyleone}%
\newtheorem{theorem}{Theorem}
\newtheorem{proposition}[theorem]{Proposition}%

\newtheorem{lemma}{Lemma}

\theoremstyle{thmstyletwo}%
\newtheorem{remark}{Remark}%

\theoremstyle{thmstylethree}%

\usepackage{amsmath,amssymb,amsthm,verbatim,mathtools,bbm,mathrsfs,float,subcaption}

\usepackage{algorithm,algpseudocode}

\usepackage{enumitem}
\usepackage{url}

\usepackage{xfrac}
\usepackage{float}
\usepackage[md]{titlesec}
\usepackage{ulem}[normalem]
\usepackage{soul,cancel}

\DeclareMathOperator{\scrA}{\mathscr{A}}

\DeclareMathOperator{\E}{\mathbb{E}}

\DeclareMathOperator{\Prob}{\mathbb{P}}

\DeclareMathOperator{\tT}{\text{T}}

\DeclareMathOperator{\tC}{\text{C}}
\DeclareMathOperator{\tG}{\text{G}}
\DeclareMathOperator{\tA}{\text{A}}

\newcommand{\abs}[1]{\left|#1\right|}

\newcommand{\ind}{\mathbbm{1}}
\newcommand{\defeq}{\coloneqq}

\newcommand{\x}{\mathbf{x}}
\newcommand{\tx}{\tilde{x}}

\newcommand{\y}{\mathbf{y}}
\newcommand{\z}{\mathbf{z}}

\newcommand{\X}{\mathcal{X}}

\newcommand{\bigO}{\mathcal{O}}
\newcommand{\bX}{\bar{X}}
\newcommand{\bY}{\bar{Y}}
\newcommand{\bA}{\mathbf{A}}
\newcommand{\bB}{\mathbf{B}}

\newcommand{\bW}{\mathbf{W}}
\newcommand{\R}{\mathbf{R}}
\newcommand{\tR}{\R}
\newcommand{\Q}{\mathbf{Q}}
\newcommand{\tQ}{\mathbf{Q}}

\newcommand{\tg}{\gamma}
\newcommand{\gmin}{\gamma_{\min}}
\newcommand{\gmax}{\gamma_{\max}}
\newcommand{\tgmin}{\tg_{\min}}
\newcommand{\tgmax}{\tg_{\max}}

\newcommand{\jrm}[1]{\textcolor{blue}{#1}}

\makeatletter
\newcommand{\mytag}[2]{%
  \text{#1}%
  \@bsphack
  \begingroup
    \@onelevel@sanitize\@currentlabelname
    \edef\@currentlabelname{%
      \expandafter\strip@period\@currentlabelname\relax.\relax\@@@%
    }%
    \protected@write\@auxout{}{%
      \string\newlabel{#2}{%
        {#1}%
        {\thepage}%
        {\@currentlabelname}%
        {\@currentHref}{}%
      }%
    }%
  \endgroup
  \@esphack
}
\makeatother
\usepackage{thmtools}

\begin{document}

\title{Posterior bounds on divergence time of two sequences under dependent-site evolutionary models}


\author[1]{\fnm{Joseph} \sur{Mathews}}\email{joseph.mathews@duke.edu}

\author*[1]{\fnm{Scott C.} \sur{Schmidler}}\email{scott.schmidler@duke.edu}

\affil[1]{\orgdiv{Department of Statistical Science}, \orgname{Duke University}, \orgaddress{\city{Durham}, \state{N.C.}, \country{USA}}}




\abstract{Let $\x$ and $\y$ be two length $n$ DNA sequences, and suppose we would like to estimate the divergence time $T$. 
Under suitable conditions, a well-known simple but crude estimate of $T$ is the fraction of differing sites $\hat{p} := \text{d}_{\text{H}}(\x,\y)/n$ (the $p$-distance).
We establish a posterior concentration bound on $T$, showing that the posterior distribution of $T$ concentrates within a logarithmic factor of $\hat{p}$ when $\text{d}_{\text{H}}(\x,\y)\log(n)/n = o(1)$.
Our bounds hold under a large class of evolutionary models and prior distributions, including many standard models that incorporate site dependence. As a special case, we show that $T$ exceeds $\hat{p}$ with vanishingly small posterior probability as $n$ increases under models with constant mutation rates, complementing the result of Mihaescu and Steel (Appl Math Lett 23(9):975–979, 2010). 
Our approach is based on bounding sequence transition probabilities in various convergence regimes of the underlying evolutionary process.  Our result is motivated by the problem of improving the efficiency of iterative optimization and sampling schemes for estimating divergence times in phylogenetic inference.}


\maketitle

\section{Introduction}
Estimation of the \textit{species divergence time} (or \textit{branch length}) $T$ between two length $n$ DNA sequences $\x$ and $\y$ is a fundamental problem in phylogenetic tree reconstruction \cite{Russo:1995, Takahata:1995, Glazko:2003,Kumar:1998,Hasegawa:1985}. Early tree reconstruction algorithms used notions of evolutionary (genetic) distance between observed sequences such as the well-known \textit{$p$-distance}, given by the proportion of differing sites $\hat{p} := \text{d}_{\text{H}}(\x,\y)/n$, with $\text{d}_{\text{H}}$ the Hamming distance, to estimate branch lengths and tree topology \cite{Fitch:1967,Fitch:1971,Saitou:1987,Rzhetsky:1992}. However, the $p$-distance underestimates divergence times when sequences are not closely related \cite{Jukes:1969,Hasegawa:1985,Kimura:1980,Yang:2008}. Later model-based approaches estimate branch lengths by assuming sequences evolve according to a continuous time Markov Chain (CTMC) with rate matrix $\tQ$ \cite{Felenstein:1973}. For such models, two popular approaches to estimation are maximum likelihood \cite{Felenstein:1981} and Bayesian inference \cite{Yang:1997,Thorne:1998}. For models with constant substitution rates (e.g. Jukes-Cantor (JC69) model \cite{Jukes:1969}), the maximum likelihood estimate of the branch length can be obtained in closed form and shown to be approximately $-\log(1-\hat{p})$. However, closed form solutions of the MLE are not available for most substitution models, and either numerical optimization or posterior sampling via Markov chain Monte Carlo (MCMC) methods are required. To ensure tractability of the likelihood, it is typically assumed that DNA sites evolve independently. Unfortunately, the independent site assumption is known to fail to capture important biological features; examples include CpG di-nucleotide mutability \cite{Pederson:1998}, structural constraints in RNA and proteins \cite{Robinson:2003}, and enzyme-driven somatic hypermutation in B-cell receptor affinity maturation \cite{Wiehe:2018,Mathews:2023b}. For this reason, a variety of \textit{dependent site} models have been proposed \cite{Robinson:2003,Siepel:2004,Jensen:2000,Larson:2020,Hwang:2004,VonHaeseler:1998,Christensen:2005,Arndt:2005,Lunter:2004}. However, under site dependent models, even numerical evaluation of the likelihood is generally intractable, further complicating the estimation of $T$. In such cases, the likelihood must be numerically approximated \cite{Robinson:2003,Jensen:2001}; doing so within iterative optimization or sampling algorithms means that estimating $T$ in dependent site settings is computationally intensive.

As a result, it is desirable to  evaluate the likelihood only at values of $T$ where the likelihood (or posterior density) is large, since likelihood approximations are expensive. Consequently, we would like to have a concentration bound on the posterior distribution of $T$, in order to identify the region containing such values. Such a bound was obtained by  \citet{Mihaescu:2010} for the  two state symmetric evolution model (CFN model), with the authors showing that $T$ exceeds $\log(n)$  with vanishingly small probability as $n$ increases. Here, we provide a much tighter bound for the setting -- common in practical applications -- of low-to-moderately mutated sequences, establishing a connection between the posterior distribution of $T$ and the $p$-distance. Specifically, we show that with high probability $T$ is no greater than $\hat{p} \log(n)$
, provided  $\hat{p}\log(n) = o(1)$. These results hold for a large class of evolutionary models, including those exhibiting site dependence. As a special case, we show that under symmetric evolution models  (e.g. 
the same setting as \cite{Mihaescu:2010}), our bound can be improved to show that $T$ is no greater than $\hat{p}$
with high posterior probability. Our results establish a direct relationship between the familiar and easily calculated $p$-distance notion of evolutionary distance and the high posterior credible region of $T$ under general evolution models including those capturing site dependence, and imply that the likelihood that $T$ lies outside this interval is exponentially small. This in turn implies that the posterior probability is exponentially small, so long as the prior is not pathological (i.e. concentrates mass outside the interval in a way that increases with data size to overwhelm the signal in the data). As a result, procedures for posterior sampling or iterative optimization of the likelihood for $T$ during phylogenetic inference need not 
consider values of $T$ outside this interval, enabling the design of more efficient proposal distributions for Bayesian MCMC, or optimization initialization procedures or constraints for maximum likelihood estimation. The latter is of particular importance for models with site dependence, where accurate approximation of the likelihood can become significantly more expensive outside this region \cite{Mathews:2025}.

This paper is organized as follows. In Section~\ref{sec: notation} we develop some notation; in Section~\ref{sec: main result} we present our main results; in Section~\ref{sec: proof of main result} we give the proofs of our main results.


\section{Notation}\label{sec: notation}
Let $\scrA$ be an alphabet (e.g. $\scrA = \{\tA,\tG,\tC,\tT\}$) of size $a := |\scrA|$ and $\x = (x_1,\ldots,x_n) \in \mathscr{A}^{n}$ be a length $n$ sequence. We assume that sequences evolve according to a time-homogeneous CTMC, with $\tg_i(b; \tx_i)$ the \textit{context-dependent} rate at which $x_i \in \scrA$ mutates to $b \in \scrA$ and
\begin{align*}
\tx_i = (x_{i},x_{c(i,1)},\ldots, x_{c(i,k)}) \in \scrA^{k+1},
\end{align*}
where $(c(i, 1), . . . , c(i, k))$ denotes the $k$ sites that belong to the context of site $i$. 
%
The case $k = 0$ corresponds to an independent site model. 
The subscript on the rates indicates a possible dependence on the site at which the mutation occurs. Note that while sequences evolve according to a time-homogeneous CTMC, the marginal process at an individual site evolves according to a time-\textit{inhomogeneous} CTMC due to the context-dependence of the rates. Let $ \tg_i(\cdot; \tx_i) = 
\sum_{b \neq x_i}  \tg_i(b; \tx_i)$ denote the rate at which site $i$ exits state $x_i$, and $\tg(\cdot ; \x) =  \sum^n_{i=1} \tg_i(\cdot; \tx_i)$ the total rate at which sequence $\x$ mutates. Denote the smallest and largest non-zero rates by
\begin{align*}
    \tgmin := \min_{i} \min_{\tilde{x}_i \in \scrA^{k+1}} \min_{b \in \scrA'(\tilde{x}_i)}   \tg_{i}(b; \tilde{x}_{i}) \quad\quad \tgmax := \max_{i} \max_{\tilde{x}_i \in \scrA^{k+1}} \max_{b \in \scrA'(\tilde{x}_i)}  \tg_{i}(b; \tilde{x}_{i}) ,
\end{align*}
where $\scrA'(\tilde{x}_i) \defeq \{b: \tg(b; \tilde{x}_i) > 0 \}$ are the set of bases reachable from $x_i$ by a single substitution. All of our results are stated under the standard assumption that multiple substitutions cannot occur simultaneously, a natural one for most sequence evolution models. 
In addition, we assume all single base substitutions are allowed at each site and context. Specifically, we assume $|\scrA'(\tilde{x}_i)| \equiv q$, where $q$ denotes the total number of single-site substitutions possible at a given site.
For example, $q = 3$ for nucleotide sequences and $q = 9$ for codon sequences. 
The sequence $\{ \x(t)\}$ evolves according to a Markov process operating over the space of all sequences with rate matrix $\tQ$ of size $a^n \times a^n$ defined by 
\begin{align}
\label{eqn: tilde Q expanded definition}
\tQ_{\x,\x^{\prime}} = 
\begin{cases}
\tg_{i}(b; \tilde{x}_{i}) & \text{ for } \text{d}_{\text{H}}(\x,\x^{\prime}) = 1 \text{ and } x^{\prime}_{i} = b \neq x_{i}  \\
-\tg(\cdot; \x) & \text{ for } \text{d}_{\text{H}}(\x,\x^{\prime}) = 0 \\
0 & \text{ for } \text{d}_{\text{H}}(\x,\x^{\prime}) > 1.
\end{cases}  
\end{align}
Let $\y \in \mathcal{A}^{n}$ be an observed descendant sequence and suppose we wish to infer the divergence time $T$ between $\x$ and $\y$. The likelihood function for $T$ is given by the transition probability from $\x$ to $\y$ under $\tQ$:
\begin{align}\label{eqn: transition probability def}
  \mathcal{L}(T \mid \x,\y) = p_{(T,\tQ)}(\y \mid \x) := (e^{T\tQ})_{\x,\y}.
\end{align}
It will be convenient to re-express \eqref{eqn: transition probability def} as a transition probability under the uniformized version of $\{\x(t) \}$, which transforms the chain to one identical in distribution but with jump times governed by a homogeneous Poisson process and state changes by a discrete time Markov chain $\tR$. (We refer the reader to Chapter 5 of \cite{Ross:1995} for details on uniformization.)
%
More precisely, note that $nq$ denotes the number of non-zero off diagonal elements in a given row of $\tQ$. Let 
$\lambda \geq \tgmax$ be any upper bound on the largest transition rate and define the probability transition matrix $\tR := \mathbf{I}_{a^{n}} + \tQ/\lambda_{\tR}$ with $\lambda_{\tR} := nq\lambda$.  Then \eqref{eqn: transition probability def} can be rewritten as 
%
\begin{align}
\label{eqn:marginal probability}
\mathcal{L}(T \mid \x,\y) = e^{-\lambda_{\tR}T} \sum^{\infty}_{m=\text{d}_{\text{H}}(\x,\y)} \frac{(\lambda_{\tR} T)^m}{m!}\tR_{\x,\y}^m .
\end{align}
Note that while setting $\lambda = \tgmax$ is valid, this choice is not necessary, and we will later have reason to choose $\lambda > \tgmax$ in proving our main results. Let $t \mapsto \eta_0(t)$ be the density of a prior distribution on $T$ and denote the corresponding posterior density of $T$ by
\begin{align*}
\eta(t \mid \x,\y) := 
\frac{\mathcal{L}(t \mid \x, \y) \eta_0(t) }{\int^{\infty}_0 \mathcal{L}(t \mid \x, \y) \eta_0(t) dt}.
\end{align*}
With a slight abuse of notation, we use the same letter for both a probability measure and its density function throughout.
\section{Main Result}
\label{sec: main result}
Our main result is the following concentration bound on $T$. The proof is given in Section~\ref{sec: proof of main result}. This bound is stated in terms of the minimum number of substitutions required to reach $\y$ from $\x$:
\begin{align*}
    r(\x,\y) = r \defeq \text{d}_{\text{H}}(\x,\y).
\end{align*}
Observe that $\x$, $\y$, and $r$ implicitly depend on $n$. Thus limits and asymptotic statements in our results are always stated as $n \rightarrow \infty$. 

%
The following result requires that the prior distribution $\eta_0$ assign sufficiently large probability to intervals of the form $J = [a \hat{p}, b \hat{p}]$, where $a,b > 0$ are model-dependent constants. Common prior distributions (e.g., $\text{Exp}(1)$) satisfy this assumption (see Remark~\ref{remark: polynomial density} below). Notice that the endpoints of $J$ scale with $\hat{p}$ (while priors are typically independent of $r$ and $n$); we discuss the motivation for this form of $J$ in Remark~\ref{remark: JC69 example} below. Essentially this condition requires that the prior not be both data-dependent and behave in such a way that it decays faster -- as a function of data size -- in the region containing the MLE than the information in the data (likelihood) grows. (That is, that it not behave pathologically).

%
\begin{theorem}
\label{thm:Posterior Probability I}
Let $J = [a \frac{r}{n} , b\frac{r}{n}]$ for constants $0 < a < b$. There exist constants $c_1 \in (0,1)$ and $c_0, c_2, c_3  \in (0,\infty)$ depending on $\gmin,\gmax, \lambda, q, a, b$, such that for $n$ sufficiently large and any two $\x,\y \in \mathcal{A}^{n}$ satisfying
\begin{align}
\label{eqn: main theorem condition}
c_0 \leq  r \leq   \frac{n}{\log(n)} c_1,
\end{align}
the following posterior tail bound on $T$ holds:
\begin{align}
\label{Eqn:TailBound}
\eta\bigg( T > c_3 \,\hat{p} \log(n) \mid \x,\y \bigg) \leq 
\frac{1}{\eta_0(J)} e^{-c_2  r \log(n)}.
\end{align}
%
In particular, if $\eta_0(t)$ has polynomial density near zero, i.e., there exists constants $C, \alpha > 0$ and $\delta \in (0,1)$ such that
\begin{align}
\label{eqn: prior polynomial density zero}
\eta_0(t) \geq C t^{\alpha} \text{ for } t \in (0,\delta),
\end{align}
then $\eta_0(J) = \Omega(n^{-(\alpha + 1)})$ and the right-hand side of \eqref{Eqn:TailBound} vanishes faster than any polynomial
in $n$ as long as $r \to  \infty$.
\end{theorem}
%
\noindent Theorem~\ref{thm:Posterior Probability I} says that for common prior distributions on $T$, the posterior probability that $T$  is larger than $\bigO(\hat{p}\log(n))$ decays super-polynomially in $n$ (faster than any power of $n$)
as long as $r \rightarrow \infty$.
As a result, procedures for posterior sampling or iterative optimization of $T$ during phylogenetic inference need not consider values of $T$ larger than $\bigO(\hat{p}\log(n))$. Note that this bound is larger than the $p$-distance, due to the $\log(n)$ term appearing in \eqref{Eqn:TailBound}; we suspect that this term is unnecessary, but have not yet been able to remove it. However, Theorem~\ref{thm: T bound symmetric} below shows that the $\log(n)$ factor can be removed for models with constant mutation rates (e.g. JC69 model). We suspect this factor can be removed for context-dependent models as well, but this appears to be non-trivial.

\begin{remark}The condition \eqref{eqn: main theorem condition} requires that $\y$ not be too highly mutated with respect to $\x$. Some such restriction that $\y$ not be too highly mutated with respect to $\x$ is necessary because when $r = \Omega(n)$, the mutational process is approaching equilibrium and the likelihood of $T$ approaches a constant (see e.g. \cite{Challis:2012})
. For example, under the JC69 model ($q=3$ and constant mutation rates) an MLE of $T$ does not exist  when $r/n \geq 3/4$. (For general models, the precise threshold at which $r/n$ becomes too large, so that the MLE no longer exists, cannot be solved for analytically.) Theorem~\ref{thm:Posterior Probability I} provides guarantees that an MLE of $T$ exists and lies in the interval $[0,\hat{p}\log(n) c)$ for a constant $c > 0$. 
\end{remark}

\begin{remark}
\label{remark: polynomial density}
Common prior distributions for the divergence time, such as the exponential and gamma distributions, satisfy \eqref{eqn: prior polynomial density zero}. For example, the exponential distribution with rate $\omega$ satisfies
\begin{align*}
\eta_0(J) = \omega \int^{b \frac{r}{n} }_{a \frac{r}{n}} e^{-\omega t} dt \geq \omega e^{-b\omega \frac{r}{n}} \frac{r}{n}\big(b - a\big) = \Omega\Big(\frac{1}{n}\Big).
\end{align*}
%
%
Moreover, the interval $J$ in Theorem~\ref{thm:Posterior Probability I} can be slightly enlarged to $J = [a \frac{r}{n} , b\log(n)\frac{r}{n}]$; the $\log(n)$ factor was omitted for readability. 
\end{remark}

\begin{remark}
\label{remark: JC69 example} 
The form of $J$ can be motivated by the following example. Consider the JC69 model with rate $\gamma \in (0,\infty)$. It can be shown that the MLE of $T$ lies inside an interval of the form $J = [a \hat{p}, b\hat{p}]$, where the interval endpoints scale with $\hat{p}$. Indeed, the MLE is
\begin{align*}
\hat{T}_{\text{JC69}} = -\frac{1}{4\gamma}\log\left(1 - \frac{4}{3} \hat{p} \right),
\end{align*}
assuming $\hat{p} < 3/4$. Using the inequality $x \leq -\log(1-x) \leq \frac{x}{1-x}$ we have 
\begin{align*}
\frac{\hat{p}}{3\gamma}   \leq \hat{T}_{\text{JC69}} \leq \frac{\hat{p}}{3\gamma(1- \frac{4}{3} \hat{p})}.
\end{align*}
Hence, we can choose $a = 1/3\gamma$. However, the right-hand side of this inequality becomes arbitrarily large as $\hat{p} \rightarrow 3/4$. The condition \eqref{eqn: main theorem condition} guarantees the existence of $c_1 \in (0,3/4)$ such that $r < n c_1$. Consequently, there exists a constant $b$ satisfying
\begin{align*}
\hat{T}_{\text{JC69}} \in [a \hat{p}, b \hat{p}], \quad \text{ for } \quad a = \frac{1}{3\gamma}, \quad b = \frac{1}{3\gamma(1- \frac{4}{3} c_{1})}.
\end{align*}
In Theorem~\ref{thm: T bound symmetric} below, we specialize Theorem~\ref{thm:Posterior Probability I} to the special case of models with constant rates.
\end{remark}

For the special case of models with constant rates, including the JC69 model, we obtain the following bound.

\begin{theorem}\label{thm: T bound symmetric}
Let $J = [a \frac{r}{n}, b \frac{r}{n}]$ for constants $0 < a < b$ as in Theorem~\ref{thm:Posterior Probability I}. Suppose that $\tQ$ corresponds to a symmetric evolution model, i.e., $\tgmax = \tgmin = \gamma \in (0,\infty)$. There exist constants $c'_0 \in (0,1)$ and $c'_1, c'_2 \in (0,\infty)$ depending on $\gamma, \lambda, q, a, b$ such
that for $n$ sufficiently large and any $\x,\y \in \mathcal{A}^{n}$ satisfying
\begin{align}
\label{eqn: main theorem sym condition}
    r \leq n c'_0 ,
\end{align}
the following posterior tail bound on $T$ holds provided $r/\log(n) \rightarrow \infty$:
\begin{align}
\label{Eqn:TailBoundSym}
    \eta\Big( T > c'_2\,  \hat{p}  \mid \x,\y \Big) \leq \frac{1}{\eta_0(J)} e^{-c'_1 r}.
\end{align}
%
%
In particular, if $\eta_0(t)$ has polynomial density near zero (see Theorem~\ref{thm:Posterior Probability I}), then the right-hand side of \eqref{Eqn:TailBoundSym} vanishes faster than any polynomial in $n$ provided $r/\log(n) \rightarrow \infty$.
\end{theorem}



\section{Proof of Theorem~\ref{thm:Posterior Probability I}}
\label{sec: proof of main result}

\subsection{Overview}
Our goal is to show that $\eta(\cdot \mid \x,\y)$ concentrates within an interval around $\hat{p} =  r/n$, the observed fraction of sites mutated. To do so, we will use the following result: 
\begin{theorem}
\label{thm:main thm appdx A}
Let $J = [a \frac{r}{n}, b\log(n) \frac{r}{n} ]$ for constants $0 < a < b$. There exist $c_1 \in (0,1)$ and constants $c_0, c_2, c_3  \in (0,\infty)$ depending on $\gmin,\gmax, \lambda, q, a,b$, such that for $n$ sufficiently large and any two $\x,\y \in \mathcal{A}^{n}$ satisfying
\begin{align}
\label{eqn: main thm appdx A r condition}
   c_0 \leq r \leq   \frac{n}{\log(n)} c_1,
\end{align}
%
%
%
the following bound holds:
\begin{align}
\frac{\sup_{T \in I^c}p_{(T,\tQ)}(\y \mid \x)}{\inf_{T^* \in J}p_{(T^*,\tQ)}(\y \mid \x)} \leq \exp(- c_2 r \log(n)) ),
\label{eqn:TRatioBound}
\end{align}
where $I := (0, c_3  \frac{r}{n}  \log(n))$.

\end{theorem}
Before proceeding to the proof of Theorem~\ref{thm:main thm appdx A}, we show that Theorem~\ref{thm:Posterior Probability I} is a direct consequence. 
\begin{proof}(Theorem~\ref{thm:Posterior Probability I})
Applying Theorem~\ref{thm:main thm appdx A} we obtain 
\begin{align*}
\eta\bigg(T > c_3 \frac{r}{n}  \log(n) \mid \x,\y\bigg) = \frac{\int_{I^c} p_{(t,\tQ)}(\y \mid \x) \eta_0(t) dt }{\int^{\infty}_0 p_{(t,\tQ)}(\y \mid \x) \eta_0(t) dt } 
&\leq \frac{\sup_{T \in I^c}p_{(T,\tQ)}(\y \mid \x) \eta_0(I^c) }{\inf_{T^{*} \in J} p_{(T^{*},\tQ)}(\y \mid \x) \eta_0(J)}  \\
&\leq \exp(-c_2 r \log(n) ) \frac{\eta_0(I^c)}{\eta_0(J)}.
\end{align*}
\end{proof}

%

%

\section{Supporting Details}

\noindent \noindent To prove Theorem~\ref{thm:main thm appdx A}, we first establish a lower bound on the denominator of \eqref{eqn:TRatioBound} by  obtaining a lower bound on \eqref{eqn:marginal probability} that holds for any $T> 0$. We do so by considering only the first ($m = r = \text{d}_{\text{H}}(\x,\y)$) term in the summation in \eqref{eqn:marginal probability} and lower bounding the probability of any length $r$ path from $\x$ to $\y$ under $\tR$.
\begin{lemma}
\label{lemma:likelihood LB}
Recall that $\lambda_{\tR} = nq\lambda$. For any $T > 0$, we have by \eqref{eqn:marginal probability}:
\begin{align*}
p_{(T,\tQ)}(\y \mid \x) \geq \exp\left(-\lambda_{\tR}T + r \log(T \tgmin) \right).
\end{align*}
In particular, for $J$ defined as in Theorem~\ref{thm:main thm appdx A},
\begin{align}
\label{eqn:LB Lemma Statement J}
    \inf_{T^* \in J}p_{(T,\tQ)}(\y \mid \x) \geq \inf_{T^* \in J} \exp\left(-\lambda_{\tR}T + r \log(T \tgmin) \right).
\end{align}
\end{lemma}
\begin{proof}
Using \eqref{eqn:marginal probability} and considering only length $r$ paths, we have
\begin{align*}
p_{(T,\tQ)}(\y \mid \x) &\geq e^{-\lambda_{\tR} T} \frac{(\lambda_{\tR}T)^r}{r!}\tR^r_{\x,\y} .
\end{align*}
Recall that $r= \text{d}_{\text{H}}(\x,\y)$, so there are $r!$ paths of length $r$ from $\x$ to $\y$. The probability of each path  is a product of $r$ off-diagonal elements of $\tR$ corresponding to the context-dependent jump probabilities. 
Each of these non-zero off-diagonal elements of $\tR$ is lower bounded by $\tg_{\min}/\lambda_{\tR}$. To see this, recall $\tR = \mathbf{I}_{a^{n}} + \tQ/\lambda_{\tR}$ and each non-zero off-diagonal element of $\tQ$ is bounded below by $\tgmin$. It follows that $\tR^{r}_{\x,\y} \geq r!  (\frac{\tg_{\min}}{\lambda_{\tR}})^r$. Hence,
\begin{align*}
e^{-\lambda_{\tR} T}\frac{(\lambda_{\tR}T)^r }{r!}\tR^r_{\x,\y} \geq \exp\left(-\lambda_{\tR}T + r \log(T \tgmin) \right).
\end{align*} 
\end{proof}
\noindent An upper bound on the numerator of \eqref{eqn:TRatioBound} is more difficult to obtain. Again, we will do so via an upper bound on \eqref{eqn:marginal probability}. In particular, we will find constants $c,c' \in (0,\infty)$ that depend on $\tgmin$, $\tgmax$, $\lambda$, and $q$, and a time $t^{*} \in (0,\infty) $ such that for $n$ sufficiently large
\begin{align}
\label{eqn:overview statement}
\sup_{T \in (t^*,\infty)}p_{(T,\tQ)}(\y \mid \x) \leq e^{- c n t^{*} },
\end{align}
%
provided that the sequences $\x$ and $\y$ satisfy $r/n \leq c'$.
Theorem~\ref{thm:main thm appdx A} then follows by combining the lower bound \eqref{eqn:LB Lemma Statement J} along
with the upper bound \eqref{eqn:overview statement} for $t^{*} = \Omega(\frac{r}{n}\log(n))$.
To bound the sum \eqref{eqn:marginal probability} we will partition the terms into three different `mixing regimes' of the jump process $\tR$, corresponding to the number of jumps (mutations). Let $p(n)$ be some polynomial in $n$ (to be chosen later) and partition the set of possible mutation counts into the three sets
%
\begin{align}
M_0(t) &\defeq \left\{r,\ldots, \lfloor  \lambda_{\tR} t /2 \rfloor \right\} \label{eqn:M sets 0} \\
M_1(t) &\defeq  \left\{\lfloor \lambda_{\tR}t /2  \rfloor+1,\ldots,p(n) \right\} \label{eqn:M sets 1} \\
M_2 &\defeq \left\{p(n) + 1,p(n)+2,\ldots \right\} \label{eqn:M sets 2},
\end{align}
%
for $\frac{2r}{\lambda_{\tR}} < t < \frac{2p(n)}{\lambda_{\tR}}$, and write \eqref{eqn:marginal probability} as
%
\begin{align}
\label{eqn:proof decomp}
p_{(T,\tQ)}(\y \mid \x) &=  e^{-\lambda_{\tR} T }\Big[ \underbrace{\sum_{m \in M_0(t) } \frac{(\lambda_{\tR} T )^m}{m!}\tR_{\x,\y}^m }_{\mytag{(i)}{eqn:First}} + \underbrace{\sum_{m \in M_1(t) } \frac{(\lambda_{\tR} T )^m}{m!}\tR_{\x,\y}^m }_{\mytag{(ii)}{eqn:Second}} \Big]  \\
&+  e^{-\lambda_{\tR} T }\underbrace{\sum_{m \in M_2 } \frac{(\lambda_{\tR} T )^m}{m!}\tR_{\x,\y}^m}_{\mytag{(iii)}{eqn:Third}} .
\end{align}
While partitioning the summation in \eqref{eqn:marginal probability} this way holds for any $\frac{2r}{\lambda_{\tR}} < t < \frac{2p(n)}{\lambda_{\tR}}$, we will specifically bound \ref{eqn:First}-\ref{eqn:Third} for the case $T \in (t,\infty)$ in order to establish \eqref{eqn:overview statement}. Let $\{X_m\}$ be a Markov chain with transition matrix $\tR$ and initial state $X_0 = \x$. Regime \ref{eqn:First} contains values of $m \geq r$ (recall $m-r$ is the number of ``extra" mutations) that are improbable under the Poisson process governing the number of jumps of $\{X_m\}$. In particular, for all $m \in M_0(t)$ we have $m < \lambda_{\tR}T$ whenever $T \in (t,\infty)$,
where $\lambda_{\tR} T$ is the mean of this Poisson process. As a result, an upper bound on \ref{eqn:First} is straightforward to obtain from Poisson tail bounds (Lemma~\ref{lemma:first likelihood UB} below) when $T \in (t,\infty)$. 

The approaches taken to bound \ref{eqn:Second} and \ref{eqn:Third} use similar but slightly different techniques.  In particular, we will bound \ref{eqn:Second} and \ref{eqn:Third} by obtaining upper bounds on the $m$ step transition probabilities $\tR^m_{\x,\y}$. We do so by analyzing the marginal process $\{\bX_m \}$ of $\{X_m \}$ defined by the Hamming distance from $\x$
\begin{align}
\label{eqn:bar X definition}
\bX_m \defeq \text{d}_{\text{H}}(X_m,\x)
\end{align}
rather than $\{X_m \}$ itself. Since $X_0 = \x$ we have $\bX_0 = 0$, and $\bX_m \in \{0,1,\ldots,n\}$ for all $m$. Recall that $\text{d}_{\text{H}}(\x,\y) = r$ and so
\begin{align}\label{eqn: Hamming distance relationship}
\tR^m_{\x,\y} \leq \sum_{\{\x^{\prime}: \text{d}_{\text{H}}(\x,\x^{\prime}) = r \}}\tR^m_{\x,\x^{\prime}} = \Prob(\bX_m = r).
\end{align}
Therefore we will be interested in bounding transition probabilities of the marginal process $\{\bX_m \}$. Our first step is to define a (discrete time) birth-death process $\{\bar{Y}_m\}$ on $\{0,\ldots,n\}$ with transition matrix $\bB$ and limiting distribution $\nu$, such that
\begin{align*}
\bX_0 = \bY_0 = 0 
\quad  \text{ and } \quad  
\bX_m \succeq \bY_m \text{ for } m = 0,1,2,\ldots 
\end{align*}
%
where $\succeq$ denotes stochastic dominance, i.e. $\Prob(\bX_m > t) \geq \Prob(\bY_m > t)$ for all $t \in \mathbb{R}$. This connection to the birth-death chain will be used to bound both \ref{eqn:Second} and \ref{eqn:Third}, in each case by bounding $\Prob(\bY_m = r)$  and then using
$\Prob(\bX_m = r)  \leq \Prob(\bY_m = r)$
to bound \eqref{eqn: Hamming distance relationship}. This approach greatly simplifies the analysis as the $\{\bY_m \}$ process is much simpler to analyze. We will bound \ref{eqn:Second} by bounding $\Prob(\bY_m = r)$ using hitting time arguments; the choice of $t$ in \eqref{eqn:M sets 1} will play an important role in this step. 
For \ref{eqn:Third}, we will bound $\Prob(\bY_m = r)$ by bounding the spectral gap of $\bB$; $p(n)$ in \eqref{eqn:M sets 2} will correspond to the resulting relaxation time bound.

\subsection{Poisson tail bound for \ref{eqn:First}}

As mentioned above, a bound for \ref{eqn:First} is straightforward to obtain. The jumps of $\{X_m\}$ are generated according to a Poisson process with mean $\lambda_{\tR}T=nTq\lambda$. Then we can apply tail probability bounds for Poisson random variables and ignore terms smaller than $\lambda_{\tR}T$.
\begin{lemma}
\label{lemma:first likelihood UB}
Let $T \in (t,\infty)$ for  $t \in (0,\infty)$. Then
\begin{align*}
e^{-\lambda_{\tR} T} \sum_{m \in M_0(t) } \frac{(\lambda_{\tR} T)^m}{m!}\tR_{\x,\y}^m 
\leq e^{-\frac{\lambda_{\tR}t }{8}}.
\end{align*}
\end{lemma}
%
%
\begin{proof}
Let $Z \sim \text{Poisson}(\lambda_{\tR}t)$.  Note that the lower tail probability of a Poisson random variable, viewed as a function of its mean, is decreasing monotonically. Since $t \leq T$, this implies
\begin{align*}
e^{-\lambda_{\tR} T} \sum_{m \in M_0 } \frac{(\lambda_{\tR} T)^m}{m!}\tR_{\x,\y}^m  \leq  e^{-\lambda_{\tR} T} \sum_{ m \leq \frac{\lambda_{\tR} t}{2} } \frac{(\lambda_{\tR} T)^m}{m!} \leq \Prob\left(Z \leq \frac{\lambda_{\tR}t}{2} \right) \leq  e^{-\frac{\lambda_{\tR}t}{8}}.
\end{align*}
The last inequality follows by the concentration bound $\Prob(Z \leq \rho (1-\epsilon)) \leq e^{-\rho h(\epsilon)}$ for $h(\epsilon) = (1-\epsilon)\log(1-\epsilon) + \epsilon$ with $\epsilon \in (0,1)$ for any Poisson random variable $Z \sim \text{Poisson}(\rho)$ (see e.g. Corollary 2.3.4 of \cite{Vershynin:2025}) with $\epsilon = 1/2$ by noting $h(\frac{1}{2}) > \frac{1}{8}$.
\end{proof}

\subsection{Coupling Construction for Bounding \ref{eqn:Second} and \ref{eqn:Third}}
\label{sec:coupling argument appdx A}
%
%
%
%
%
We first describe the time evolution of \eqref{eqn:bar X definition}. $\bX_m$ decreases by one if $X_m$ transitions to a sequence in the  (random) set:
\begin{align*}
S^{-}_m(X_m) = S^{-}_m \defeq \{\x^{\prime}: \text{d}_{\text{H}}(\x^{\prime},\x) = \bX_m-1 \text{ and } \text{d}_{\text{H}}(\x^{\prime},X_m) = 1 \},
\end{align*}
which is of size $\abs{S^{-}_m} = \bX_m$ since there are $\bX_m = \text{d}_{\text{H}}(X_m,\x)$ possible sites at which to increase the agreement with $\x$. $\bX_m$ remains unchanged if $X_m$ transitions to a sequence in the set 
\begin{align*}
S^0_m(X_m) = S^0_m \defeq \{\x^{\prime}: \text{d}_{\text{H}}(\x^{\prime},\x) = \bX_m \text{ and } \text{d}_{\text{H}}(\x^{\prime},X_m) \leq 1 \},
\end{align*}
which is of size $\abs{S^0_m} = (q-1)\bX_m + 1$
accounting for 
$(q-1)$ possible mutations at each of the $\bX_m$ differing sites, plus the possibility of $X_m$ transitioning to itself (a `virtual jump' in the uniformized process). Let $S_m(X_m) \defeq \{\x^{\prime}: \text{d}_{\text{H}}(\x^{\prime},X_m) \leq 1  \}$ denote the set of sequences corresponding to the positive elements of $\tR$ along the row corresponding to $X_m$. $\bX_m$ increases by one if $X_m$ transitions to a sequence in the set
\begin{align*}
S^{+}_m(X_m) = S_m^{+} := S_m(X_m) \setminus (S_m^{-}(X_m) \cup S^0_m(X_m)).
\end{align*}
Note $\abs{S_m} = nq + 1$. Therefore
\begin{align}
\label{eqn:BD Cards}
\abs{S_m^{-} \cup S^0_m} = q\bX_m + 1 \qquad \text{ and }\qquad \abs{S_m^{+}} = q(n-\bX_m).
\end{align}
Although $\{\bX_m\}$ is not a Markov process, we will relate it to a birth-death process. Call a transition a `birth' if $\bX_{m+1} = \bX_m+1$ and a `death' if $\bX_{m+1} = \bX_m -1$. The corresponding birth and death probabilities are given by: 
%
\begin{align}
&b(X_m) \defeq \sum_{\x' \in S_m^{+}} \frac{\tg_{i_m(\x')}(x_{i_m(\x')}; \tilde{x}_{i_m(\x')})}{n q \lambda } 
\geq \frac{\tgmin}{\lambda}\left(1- \frac{\bX_m}{n}\right) =:  b^{\prime}(\bX_m)\label{eqn:BD probs} \\
&d(X_m) \defeq \sum_{ \x' \in S^{-}_m    } \frac{\tg_{i_m(\x')}(x_{i_m(\x')}; \tilde{x}_{i_m(\x')})}{n q \lambda } \leq \frac{\tgmax \bX_m}{nq\lambda} =: d^{\prime}(\bX_m) \label{eqn:BD probs 2},
\end{align}
where $i(\x',X_m) = i_m(\x')$ denotes the site where $\x'$ and $X_m$ differ, and $b'$, $d'$ represent uniform bounds. Conditional on the \textit{sequence} $X_{m-1}$, the one step transition probability for the \textit{distance} $\bX_m$ is
\begin{align}
\label{eqn:bar x process}
\Prob(\bX_m = j \mid X_{m-1}) \defeq
\begin{cases} 
 d(X_{m-1}) & \text{ if } j = \bX_{m-1}-1 \\
1 - d(X_{m-1})  - b(X_{m-1})  & \text{ if } j = \bX_{m-1} \\
b(X_{m-1})   & \text{ if } j = \bX_{m-1} + 1\\
0 & \text{ otherwise. }
\end{cases}
\end{align}
At times, we let $\bA(X_{m-1}, \cdot)$ denote the $a^n\times (n+1)$ transition probability matrix \eqref{eqn:bar x process}.

\vspace{0.5em}
\textit{Example: Consider the JC69 model, so $q=3$ and rates are constant $\tg_{i}(b; \tilde{x}) \equiv \gamma \in (0,\infty)$, and let $\lambda > \gamma$. Then $\{\bX_m \}$ is a Markov chain and forms a true birth-death process, since the off-diagonal elements of $\tR$ are identical, and so}
\begin{align}
\label{eqn:bar x process simple}
\Prob(\bX_m = j \mid \bX_{m-1} = i)  =
\begin{cases}
\frac{i \gamma}{3n \lambda} & \text{ if }  j=i-1 \\
1 - \frac{\gamma}{\lambda} +  \frac{2i \gamma}{3n \lambda } & \text{ if }  j=i \\
\frac{\gamma}{\lambda}(1 - \frac{i}{n}) & \text{ if }  j=i+1 \\
0 & \text{ otherwise. }
\end{cases}
\end{align}
\vspace{0.5em}
%


We will bound the transition probabilities $\Prob(\bX_m = r \mid \bX_0 = 0)$ for the marginal process $\{\bX_m\}$ in terms of a corresponding birth-death process of the form \eqref{eqn:bar x process simple} constructed by `minorizing' $\{\bX_m\}$. In particular, we will define a birth-death process $\{\bY_m \}$ such that 
%
\begin{align}
\Prob(\bX_m = r \mid \bX_0 = 0) \leq 
\Prob(\bX_m \leq r \mid \bX_0 = 0) \leq 
\Prob(\bY_m \leq r \mid \bY_0 = 0),
\label{Eqn:BDbound}
\end{align}
%
holds for all $m=0,1,2,\ldots$. Notice that this will hold for any birth-death process $\{\bY_m\}$ having both a smaller birth probability and a higher death probability than the $\{\bX_m \}$ process. Such a minorizing birth-death chain can therefore be obtained by assigning birth and death probabilities given by the uniform lower and upper bounds $b^{\prime}(\bX_m)$ and $d^{\prime}(\bX_m)$ on $b(X_m)$ and $d(X_m)$ given in \eqref{eqn:BD probs}, which  depend only on $\bar{X}_m$ as in \eqref{eqn:bar x process simple}. 
We will proceed to bound the transition probabilities of the marginal process $\{\bX_m\}$ in terms of those of $\{\bY_m\}$ as in \eqref{Eqn:BDbound}.

Before stating the result, we note that if $X \succeq Y$ for $X \sim \mu$ and $Y \sim \nu$, then there exists a coupling $(X^*,Y^*)$ of $\mu$ and $\nu$ such that $X^* \sim \mu$, $Y^* \sim \nu$, and $Y^* \leq X^*$ with probability one (see e.g. \cite{Ross:1995} Ch. 9). 
%
%

%
\begin{lemma}
\label{lemma:BD process minor}
Let $\{X_m \}$ be a Markov chain with transition matrix $\tR$ with $X_0 = \x$ and $\{\bX_m \}$ its marginal process \eqref{eqn:bar x process}. Let $\lambda \geq \max(\tgmax,\tgmax q^{-1} + \tgmin) $
and consider the birth-death process $\{\bY_m \}$ with transition matrix $\bB$ given by 
\begin{align}
\label{eqn:BD process minor}
\bB_{ij} \defeq  \Prob(\bY_m = j \mid \bY_{m-1} = i) =
\begin{cases}
\frac{\tgmax i}{nq\lambda} & \text{ if } j = i-1 \\
1-\frac{\tgmax i}{nq\lambda}  - \frac{\tgmin}{\lambda}\left(1- \frac{i}{n}\right)  & \text{ if } j = i \\
\frac{\tgmin}{\lambda}\left(1- \frac{i}{n}\right)  & \text{ if }  j = i+1 \\
0 & \text{ otherwise. }
\end{cases}
\end{align}
Then  $\Prob(\bX_m \leq r \mid \bX_0 = 0) \leq \Prob(\bY_m \leq r \mid \bY_0 = 0)$ and as a consequence
\begin{align*}
\tR_{\x,\y}^m \leq \Prob(\bX_m \leq r\mid \bX_0 = 0) \leq \Prob(\bY_m \leq r  \mid \bY_0 = 0).
\end{align*}
%
\end{lemma}

\begin{proof}
Let $b^{\prime}(\bY_m)$ and $d^{\prime}(\bY_m)$ be the birth and death probabilities defined in \eqref{eqn:BD probs} and used in \eqref{eqn:BD process minor}. To bound the margin $\{\bX_m\}$ of $\{X_m\}$ in terms of $\{\bY_m\}$, we will define a coupling $(X^*_0,\ldots,X^*_m,\bY^*_0,\ldots,\bY^*_m)$ satisfying
\begin{align}
\Prob(X^*_j = \x_j \mid X^*_{j-1} = \x_{j-1}) = \tR_{\x_{j-1},\x_j} \
\quad   \text{ and }  \quad \Prob(\bY^*_j = y_j \mid \bY^*_{j-1}  = y_{j-1}) = \bB_{y_{j-1},y_{j}}
\label{Eqn:CouplingMargins}
\end{align}
with $\bar{Y}^*_0 = \bar{Y}_0 = \bar{X}^*_0 = \bar{X}_0 = 0$ and such that $\Prob(\cap^m_{j=0}\{\bX^*_j \geq \bY^*_j\}) = 1$. In words, \eqref{Eqn:CouplingMargins} says that the $\{X^*_j\}$ and $\{\bY^*_j \}$ processes are Markov chains with transition matrices $\tR$ and $\mathbf{B}$, respectively. Note that \eqref{Eqn:CouplingMargins} also implies $\mathcal{L}(\bX_1,\ldots,\bX_m) = \mathcal{L}(\bX_1^*,\ldots,\bX_m^*)$ for $\bX_j^* \defeq \text{d}_{\text{H}}(X^*_j,\x)$. The coupling construction will proceed inductively by setting $(X_0^*,\bY_0^*) = (\x, 0)$ (so $\bX_0^* = 0$) and constructing $(X_j^*,\bY_j^*)$ conditional on $(X_{i<j}^*,\bY_{i<j}^*)$ for each subsequent time step $j=1,\ldots,m$. At each step we first construct the coupling $(\bX^*_j,\bY^*_j)$ conditional on $(X^*_{i<j},\bY^*_{i<j})$ between $\bar{Y}_j$ and the marginal $\bX_j$ of $X_j$, and then construct $X^*_j  \mid \bX^*_j$ conditionally to complete the construction of the coupling $\{X^*_0,\ldots,X^*_j,\bY^*_0,\ldots,\bY^*_j\}$.  Specifically, the coupled random processes can be simulated as follows:
\begin{enumerate}
\item[(a)] Draw $\bX^*_j \mid X^*_{j-1}  \sim \bA(X^*_{j-1}, \cdot)$ and $\bY^*_j  \mid \bY^*_{j-1} \sim \bB$ such that $\Prob(\bar{X}^*_j \geq \bar{Y}^*_j) = 1$ (see below).
\item[(b)] Draw $X_j^* \mid \bX^*_j, X^*_{j-1}$ from distribution
\begin{equation*}
\Pr(X^*_j = \x^{\prime}\mid \bX^*_j, X^*_{j-1}) =
\begin{cases}
\frac{\tR_{X^*_{j-1},\x^{\prime}}}{\sum_{\x^{\prime \prime}}\tR_{X^*_{j-1},\x^{\prime \prime}}\ind_{\bX^*_j}(\text{d}_{\text{H}}(\x^{\prime \prime},\x))} & \text{ if }  \text{d}_{\text{H}(\x^{\prime},\x)} =  \bX^*_j\\
 0 & \text{ otherwise }
\end{cases}
\end{equation*}
\end{enumerate}
Marginalizing out $\bX^*_j$ we see that $X^*_j \mid X^*_{j-1} \sim \tR_{X^*_{j-1}, \cdot}$  since by definition 
\begin{align*}
\Prob(X_j^* = \x^{\prime} \mid X^*_{j-1}) & =
\sum^n_{k=0} \Prob(\bX_j^* = k \mid X_{j-1}^*) \Prob(X_j^* = \x' \mid \bX_j^*, X_{j-1}^*)\\
& = \sum^{n}_{k=0} 
\bA(X^*_{j-1},k) 
\frac{\tR_{X^*_{j-1},\x^{\prime}} \ind_{k}(\text{d}_{\text{H}}(\x^{\prime},\x)) }{\sum_{\x^{\prime \prime}}\tR_{X^*_{j-1},\x^{\prime \prime}}\ind_{k}(\text{d}_{\text{H}}(\x^{\prime \prime},\x))} \\ 
&= \sum^{n}_{k=0} \tR_{X^*_{j-1},\x^{\prime}} \ind_{k}(\text{d}_{\text{H}}(\x^{\prime},\x)) 
= \tR_{X^*_{j-1},\x^{\prime}}. 
\end{align*}
To show that step $(a)$ is possible, we give an explicit construction for generating $\bar{X}_j^*$ and $\bar{Y}_j^*$ that satisfy these properties. To do so, we will show stochastic dominance holds for the conditional distributions
\begin{equation*}
\bX_j\mid X_{j-1} \succeq \bY_j \mid \bY_{j-1}
\end{equation*}
whenever $\bX_{j-1} \succeq \bY_{j-1}$. That is, we require to show that for $t \in \{0,\ldots,n\}$
\begin{align}
\label{eqn:step}
\Prob\left(\bX_j > t \mid X_{j-1}, \bar{X}_{j-1} = x   \right) - \Prob\left(\bY_j > t \mid \bY_{j-1} = y  \right)  \geq 0 \qquad \text{ if } \quad x \geq y.
\end{align}
We consider the cases $x > y$ and $x = y$ separately. 
\\
\\
\textit{Case $x > y$}: Fix $t \in \{0,\ldots,n\}$. For $y \in \{0,\ldots,t-1\}$, $\Prob\left(\bY_j > t \mid \bY_{j-1} = y  \right) = 0$ and so \eqref{eqn:step} holds. Similarly, if $y \in \{t+1,\ldots,n \}$, then $x \in \{t+2,\ldots,n\}$, so $\Prob\left(\bX_j > t \mid X_{j-1}, \bar{X}_{j-1} = x   \right)  = 1$ in which case \eqref{eqn:step} holds. Thus the only nontrivial situation occurs when $y = x - 1 = t$. Then we have 
\begin{align*}
\Prob\left(\bX_j > t  \mid X_{j-1}, \bX_{j-1} = t+1  \right) - \Prob\left(\bY_j > t \mid \bY_{j-1} = t  \right) 
& = 1-d(X_{j-1}) - b^{\prime}(t) \\
&\geq 1 - d^{\prime}(t+1) - b^{\prime}(t) \\
&= 1 - \frac{\tgmax(t+1)}{nq\lambda} - \frac{\tgmin}{\lambda}\left(1 - \frac{t}{n} \right) \\
&\geq 0.
\end{align*}
where the first inequality follows from
\eqref{eqn:BD probs 2} and the final inequality follows by our choice of $\lambda$. \\
It follows that for $x > y$:
\begin{align*}
\Prob\left(\bX_j > t  \mid X_{j-1}, \bar{X}_{j-1} = x  \right) - \Prob\left(\bY_j > t \mid \bY_{j-1} = y  \right) > 0.
\end{align*}
\\
\textit{Case $x = y$:} First consider $x=y=t+1$
. Then we have 
\begin{align*}
\Prob\left(\bX_j > t  \mid X_{j-1}, \bX_{j-1} = t+1 \right) - \Prob\left(\bY_j > t \mid \bY_{j-1} = t+1  \right) 
&= 1 - d(X_{j-1}) - (1 - d^{\prime}(t+1)) \\
&\geq 0.
\end{align*}
The only other non-trivial case to consider is $x = y = t$, which from \eqref{eqn:BD probs} we obtain 
\begin{align*}
\Prob\left(\bX_j > t  \mid X_{j-1}, \bX_{j-1} = t  \right) - \Prob\left(\bY_j > t \mid \bY_{j-1} = t  \right) = b(X_{j-1}) - b^{\prime}(t) \geq 0.
\end{align*}
Consequently, conditional on $x=y$
\begin{align*}
\Prob\left(\bX_j > t  \mid X_{j-1}, \bX_{j-1} = x \right) - \Prob\left(\bY_j > t \mid \bY_{j-1}  = y  \right)\geq 0.
\end{align*}
This establishes the second case.

Since \eqref{eqn:step} holds in both cases, it follows that there exists a coupling $(\bX_j^*,\bY_j^*)$ of $\bA(X^*_{j-1},\cdot)$ and $\bB_{\bY^*_{j-1}, \cdot}$ such that $\Prob(\bX_j^* \geq \bY_j^* \mid \bX^*_{j-1},\bY^*_{j-1}) = 1$.  Hence, $\Prob(\cap^m_{j=0} \{\bX^*_j  \geq \bY^*_j \} ) = 1$ for any $m$ and, in particular,
 \begin{align*}
\Prob(\bar{X}_m \leq r \mid X_0 = \x) = \Prob(\bar{X}^*_m  \leq r \mid X^*_0 = \x) \leq \Prob(\bar{Y}^*_m \leq r \mid \bY^*_0 = 0) = \Prob(\bar{Y}_m \leq r \mid \bar{Y}_0 = 0).
 \end{align*}
\end{proof}
Note that  Lemma~\ref{lemma:BD process minor} is tight  in the case $\tgmax = \tgmin$ and $q = 3$ in \eqref{eqn:bar x process simple}. By Lemma~\ref{lemma:BD process minor}, it suffices to analyze the transition matrix $\bB$ instead of $\tR$ in order to bound \ref{eqn:Second} and \ref{eqn:Third} in \eqref{eqn:proof decomp}.

\subsection{Bounding \ref{eqn:Second} Using Hitting Times of $\bY_m$}
Recall that \eqref{eqn: Hamming distance relationship} implies that a bound on \ref{eqn:Second} can be obtained by bounding $\Prob(\bX_m = r \mid \bX_0 = 0)$. However, by Lemma~\ref{lemma:BD process minor} there is a birth-death process $\{\bY_m\}$ with $\bY_0 = 0$ and transition probabilities \eqref{eqn:BD process minor} such that
\begin{align*}
\Prob(\bX_m = r \mid \bX_0 = 0) \leq \Prob(\bX_m \leq r \mid \bX_0 = 0) \leq \Prob(\bY_m \leq r \mid \bY_0 = 0).
\end{align*}
Consequently, it suffices to bound $\Prob(\bY_m \leq r \mid \bY_0 = 0)$ for all $m \in M_1(t)$ in order to bound \ref{eqn:Second}. For brevity we will write $\Prob(\bY_m \leq r \mid \bY_0 = 0) = \Prob_0(\bY_m \leq r)$, with the subscript denoting the initial state. 

Let $r^* > r$ be some reserved state (to be chosen later), and $\tau_{r^*} \defeq \inf\{m^{\prime}: \bY_{m^{\prime}} = r^* \}$ its associated hitting time.  We will bound $\Prob_0(\bY_m \leq r)$ by considering the cases $m \leq \tau_{r^*}$ and 
$m > \tau_{r^*}$, via
\begin{align}
\Prob_0(\bY_m \leq r) \leq \Prob_0(\tau_{r^*} > m) + \Prob_0(\bY_m \leq r, \tau_{r^*} \leq m).
\label{Eqn:Yrdecomp}
\end{align}
%

%
%
The state $r^*$ will be chosen later to be a `point of no return' for state $r$, in the sense that conditional on the event $\{\tau_{r^{*}} \leq m\}$, the probability that $\bY_m$ is less than $r$ decays exponentially in the sequence length $n$, allowing us to obtain an upper bound on the second term $\Prob_0(\bY_m \leq r, \tau_{r^*} \leq m)$ in \eqref{Eqn:Yrdecomp}. We will bound the first term $\Prob_0(\tau_{r^*} > m)$ by `minorizing' $\{ \bY_m\}$ by a pure random walk process $\{\bar{W}_m\}$, following a similar argument to that used to bound $\{\bX_m \}$ by $\{\bY_m\}$ in Lemma~\ref{lemma:BD process minor},
and then bounding the hitting time for $r^*$ under the random walk  $\{\bar{W}_m\}$. 
\begin{lemma}
\label{lemma:first hit BD process}
Let $r^* = \lfloor n\delta \rfloor$ with $\delta \leq \frac{tq\tgmin}{2t (\tgmax + q \tgmin}) + 4 =: \delta_1(t) = \delta_1$, and let $m^* \in M_1(t)$. Then $\Prob_0(\tau_{r^*} > m^*) \leq \exp(-\frac{m^*\tg^2_{\min}}{8 \lambda^2} )$.
\end{lemma}
%
\begin{proof}
Consider a biased random walk $\{\bar{W}_m \}$ on $\mathbb{Z}$ with $\bar{W}_0 = 0$ with transition probabilities
\begin{align}
\label{eqn:Walk}
\bW_{i,j} = \Prob(\bar{W}_m = j \mid \bar{W}_{m-1} = i) =
\begin{cases}
\frac{\tgmax \delta}{q\lambda} & \text{ if } j = i-1 \\
1-\frac{\tgmax \delta}{q\lambda}  - \frac{\tgmin}{\lambda}\left(1-\delta\right)  & \text{ if } j = i \\
\frac{\tgmin}{\lambda}\left(1- \delta\right)  & \text{ if }  j = i+1 \\
0 & \text{ otherwise. }
\end{cases}
\end{align}
given by uniform bounds on the transition probabilities of $\{ \bY_m\}$ in \eqref{eqn:BD process minor} over the range of values $0,\ldots,r^*$:
\begin{align*}
    \bW_{i,i-1} \geq \bB_{i,i-1} \qquad \bW_{i,i+1} \leq \bB_{i,i+1}.
\end{align*}
Indeed, since $r^* = \lfloor n\delta \rfloor$, the birth and death probabilities of $\{\bY_m\}$ and $\{\bar{W}_m \}$ satisfy 
\begin{align}
\bB_{i-1,i} &= \frac{\tgmin}{\lambda}\left(1- \frac{(i-1)}{n}\right)  \geq \frac{\tgmin}{\lambda}(1-\delta) = \bW_{i-1,i}, \label{eqn:walk bound 1} \\
\bB_{i,i-1} & = \frac{\tgmax i}{nq\lambda}   \leq \frac{\tgmax \delta}{q\lambda} = \bW_{i,i-1} \qquad \text{ for all } i \in \{1,\ldots,r^*\} \label{eqn:walk bound 2}.
\end{align}
Define $\eta_{r^*} := \inf\{m^{\prime}: \bar{W}_{m^{\prime}} = r^*  \}$ and the stopped processes
\begin{align*}
  \bY_{m \wedge \tau_{r^*}}   = 
    \begin{cases}
        \bY_m &\text{ for } m < \tau_{r^*} \\
        r^* &\text{ for } m \geq \tau_{r^*}
    \end{cases}  \quad\quad  
 \bar{W}_{m \wedge \eta_{r^*}} =   \begin{cases}
        \bar{W}_m &\text{ for } m < \eta_{r^*} \\
        r^* &\text{ for } m \geq \eta_{r^*}.
    \end{cases}
\end{align*}
%
%
%
By \eqref{eqn:walk bound 1} and \eqref{eqn:walk bound 2}, and recalling that $\bY_{0 \wedge \tau_{r^*}} = \bY_0 = 0$ and $\bar{W}_{0 \wedge \eta_{r^*}} = \bar{W}_0 = 0$, a stochastic dominance argument near identical to the one in the proof of Lemma~\ref{lemma:BD process minor} implies that we can define a coupling
$\{(\bY^{\prime}_{m \wedge \tau_{r^*}}, \bar{W}^{\prime}_{m \wedge \eta_{r^*}})\}$ on $\{0,\ldots,r^{*}\} \times \mathbb{Z}$ such that $\bY^{\prime}_{m \wedge \tau_{r^*}} \succeq \bar{W}^{\prime}_{m \wedge \eta_{r^*}}$, which implies
\begin{align}
\label{eqn:hitting time random walk 2nd eq}
\Prob_0(\tau_{r^*} > m^*) \leq \Prob_0(\eta_{r^*} > m^*). 
\end{align}
%
By \eqref{eqn:hitting time random walk 2nd eq}, it suffices to bound $\Prob_0(\eta_{r^*} > m^*)$. To bound this term, let $\Delta_1,\ldots,\Delta_m$ be the sequence of $i.i.d.$ increments $\Delta_i = \bar{W}_i - \bar{W}_{i-1}$ with
\begin{align*}
\Prob(\Delta_i = j) =
\begin{cases}
\frac{\tgmax \delta}{q\lambda} & \text{ if } j = -1 \\
1-\frac{\tgmax \delta}{q\lambda}  - \frac{\tgmin}{\lambda}\left(1-\delta\right)  & \text{ if } j = 0 \\
\frac{\tgmin}{\lambda}\left(1- \delta\right)  & \text{ if }  j = 1 \\
0 & \text{ otherwise. }
\end{cases}
\end{align*}
Observe that if $\sum^{m^*}_{i=1} \Delta_i > r^*$ then $\eta_{r^*} \leq m^*$, and so
$\Prob_0(\sum^{m^*}_{i=1} \Delta_i > r^*) \leq \Prob_0(\eta_{r^*} \leq m^*)$.
Let $\rho := \E[\Delta_1]  = \frac{\tgmin}{\lambda}  - \delta(\frac{\tgmax}{q\lambda} + \frac{\tgmin}{\lambda})$. Then since $r^{*} = \lfloor n\delta \rfloor$, and $m^* > n\lambda tq / 2$ (since $m^* \in M_1(t)$), we have:
\begin{align*}
\Prob_0(\eta_{r^*} > m^*) \leq \Prob\left( \frac{1}{m^*}\sum^{m^*}_{i=1}\Delta_i - \rho \leq \frac{n\delta}{m^*} - \rho  \right) &\leq  \Prob\left( \frac{1}{m^*}\sum^{m^*}_{i=1}\Delta_{i} - \rho \leq \frac{2\delta}{\lambda t q} - \rho \right).
\end{align*}
By our choice of $\delta$ we have that $\frac{2\delta}{\lambda t q} - \rho \leq - \frac{\tgmin}{2 \lambda}$. The stated bound follows by Hoeffding's inequality.
  \end{proof}
%
%

%

%

Having dealt with the term $\Prob_0(\tau_{r^*} > m)$ in \eqref{Eqn:Yrdecomp} via  Lemma~\ref{lemma:first hit BD process}, we now turn to bounding the term $\Prob_0(\bY_m \leq r, \tau_{r^*} \leq m)$, the probability that $\bY_m \leq r$ given that $\bY$ has already passed through $r^*$ (recall $r^* > r$). Denote by $\nu$ the stationary distribution of the birth-death process $\{\bY_m\}$ with transition probability matrix $\bB$. We now show that $\nu$ is a binomial distribution, and state some additional facts which we will often appeal to later in our analysis.
\begin{lemma}
\label{lemma:BD stationary}
Let $\tg_* := \tgmax/\tgmin$ be the ratio of the largest to smallest transition rates. The stationary distribution of $\bB$, denoted $\nu(x)$, is Binomial$(n,q/(q+\tg_*))$.
In addition, the following bounds hold: 
\begin{enumerate}
\item $\min_x\nu(x) = \frac{\min(\tg^{n}_{*},q^{n})}{(\tg_*}+q)^{n}$.
\item $\sum^r_{k=0}\nu(k) \leq e^{-\frac{n}{2}(\frac{q}{\tg_* + q})^2}$ provided $\frac{r}{n} < \frac{q}{2(\tg_* + q )}$.
\end{enumerate}
%
\end{lemma}
Note that the second condition in Lemma~\ref{lemma:BD stationary} requires that $r$ be less than half the average number of mutated sites under $\nu$, meaning that $r$ is a low probability state under $\nu$ by Hoeffding's inequality. 

%
\begin{proof}
Recall the definition of $\bB$ (Lemma~\ref{lemma:BD process minor}) 
\begin{align}
\label{eqn:transition probs}
\bB_{i,i-1} = \frac{\tgmax i}{nq\lambda} \quad\quad \bB_{i-1,i} = \frac{\tgmin}{\lambda}\left(1 - \frac{i-1}{n}\right).
\end{align}
Recall $\nu(x) \propto \prod^x_{i=1} \bB_{i-1,i}/\bB_{i,i-1}$ and notice
\begin{align*}
\prod^x_{i=1} \frac{\bB_{i-1,i}}{\bB_{i,i-1}} = \theta^x\prod^x_{i=1} \left(\frac{n-i+1}{i}\right) = \theta^x \binom{n}{x}.
\end{align*}
where $\theta := q\tgmin/\tgmax = q/\tg_*$. Hence, $\nu(x) = \frac{\theta^x \binom{n}{x}}{(1+\theta )^{n}}$. The minimum probability is obtained at one of the extrema:
\begin{align*}
\frac{\theta^x \binom{n}{x}}{(1+\theta )^{n}} \geq \frac{\min(1,\theta^{n})}{(1+\theta)^{n}} = \frac{\min(\tg^n_*,q^n)}{(\tg_* + q)^n}  = \min_x \nu(x).
\end{align*}
Finally, let $X \sim \text{Binomial}( n,\frac{\theta}{1+\theta})$. By Hoeffding's inequality and our assumption $r/n < \frac{q}{2(\tg_* + q)} = \frac{\theta}{2(1 + \theta)}$,
\begin{align*}
\sum^r_{k=0}\nu(k) = \Prob(X \leq r) &= \Prob\left(X-\frac{n\theta}{1+\theta} \leq -n \left(\frac{\theta}{1+\theta} -\frac{r}{n}\right) \right) \\
&\leq \Prob\left(X-\E[X] \leq - \frac{nq}{2(\tg_* + q)} \right) \leq e^{-\frac{n}{2}(\frac{q}{\tg_* + q})^2}.
\end{align*}
\end{proof}

We now bound $\Prob_0(\bY_m \leq r, \tau_{r^*} \leq m)$ using the following bound  for discrete time Markov processes on finite spaces 
, which bounds the probability of hitting $r$ before returning to $r^*$ when starting from $r^*$. We defer the proof of this result to the appendix, which follows directly by existing results given in \cite{Levin:2009}.
\begin{restatable}{lemma}{hittingtime}\label{lemma:return time bound}
Let $\{Z_t \}$ be an ergodic discrete time Markov chain defined on a finite state space $\mathcal{X}$ with stationary distribution $\nu$. For $i \in \mathcal{X}$, define the hitting times
\begin{align*}
\tau^{+}_i := \inf\{t \geq 1: Z_t = i \} \quad \text{ and } \quad \tau_i := \inf\{t \geq 0: Z_t = i \}.
\end{align*}
Then for $r,r^* \in \mathcal{X}$, the following bound holds:
\begin{align*}
    \Prob(\tau_{r} < \tau^{+}_{r^*} \mid Z_0 = r^*) \leq \frac{\nu(r)}{\nu(r^*)}.
\end{align*}
\end{restatable}
%
\noindent Lemma~\ref{lemma:return time bound} gives us the following bound:
%
\begin{align}
\label{eqn:metastable bound}
\Prob(\bar{Y}_m = r, \bar{Y}_{m - 1} \neq r^{*},\ldots,\bar{Y}_{m^{\prime}+1} \neq  r^{*} \mid \bY_{m^{\prime}} = r^*) \leq \frac{\nu(r)}{\nu(r^*)} \text{ for } m^{\prime} < m.
\end{align}
%
Lemma~\ref{lemma:first hit BD process} bounds the $\Prob_0(\tau_{r^*} > m)$ term appearing in \eqref{Eqn:Yrdecomp}, corresponding to the `escape time' for the chain to hit the state $r^*$. We will use \eqref{eqn:metastable bound} to show that the term $\Prob_0(\bY_m \leq r, \tau_{r^*} \leq m)$ appearing in \eqref{Eqn:Yrdecomp} is bounded above by $(r+1)m e^{-\frac{n\delta}{2}+1}$, and therefore decays exponentially in $n$ for $m \in M_1(t)$
for appropriately chosen $\delta$ and $r$ (and hence $r^*$). We will use the following lemma to show this.
\begin{lemma}
\label{lemma:stationary bound BD process}
Let $ \delta \leq \frac{q}{q + e\tg_*} =: \delta_2$, where we recall $\tg_* = \tgmax/\tgmin$, and suppose $r^* = \lfloor n\delta  \rfloor$.
Then 
$\max_{j \in \{0,\ldots,r\} }\frac{\nu(j)}{\nu(r^*)} \leq e^{-\frac{n\delta}{2}+1}$ for $r \leq \lfloor n \delta \rfloor/2$.
\end{lemma}
\begin{proof}
Recall that $\nu(x) \propto \prod^x_{i=1} \bB_{i-1,i}/\bB_{i,i-1}$. By the definition of $\bB$ (Lemma~\ref{lemma:BD process minor}) 
\begin{align*}
\bB_{i-1,i} = \frac{\tgmin}{\lambda}\left(1- \frac{i-1}{n}\right)  \quad\quad \bB_{i,i-1} = \frac{\tgmax i}{nq\lambda}  
\end{align*}
Observe that for any $i \in \{r+1,\ldots,r^{*}\}$ we have since $r^* = \lfloor n \delta \rfloor \leq n\delta$,  
\begin{align*}
\frac{\bB_{i,i-1}}{\bB_{i-1,i}} = \frac{\tg_*}{q}\left(\frac{i}{n - i + 1} \right) \leq \frac{\tg_*}{q}\left(\frac{r^{*}}{n - r^{*}} \right) \leq \frac{\tg_*}{q}\left( \frac{\delta}{1-\delta}\right).
\end{align*}
Note that $\frac{\tg_*}{q}( \frac{\delta}{1-\delta}) \leq e^{-1}$ by our choice of $\delta$. In addition, note that  $r^{*} - r = \lfloor n \delta \rfloor - r \geq  \lfloor n \delta \rfloor - \frac{\lfloor n \delta \rfloor}{2} = \frac{\lfloor n \delta \rfloor}{2}$. Thus we obtain for any $j \in \{0,\ldots,r\}$ 
\begin{align*}
\frac{\nu(j)}{\nu(r^*)} = \prod^{r^*}_{i=j+1} \frac{\bB_{i,i-1}}{\bB_{i-1,i}} \leq \left(\frac{\tg_*}{q} \left(\frac{\delta}{1-\delta}\right) \right)^{r^* - r} \leq e^{-\frac{\lfloor n\delta \rfloor }{2}} \leq e^{-\frac{n\delta}{2} + 1},
\end{align*}
where the final inequality follows since $\lfloor x \rfloor \geq x-1$. Since this inequality holds for any $j \in \{0,\ldots,r\}$, we obtain the stated bound.
\end{proof}

We will now use the inequality \eqref{eqn:metastable bound} along with Lemma~\ref{lemma:stationary bound BD process} to bound the term $\Prob_0(\bY_m \leq r, \tau_{r^*} \leq m)$ in \eqref{Eqn:Yrdecomp}.

%
\begin{lemma}
\label{lemma:return time BD process}
Let $\delta$, $r$, and $r^*$ satisfy the same assumptions as in Lemma~\ref{lemma:stationary bound BD process}. Then the following bound holds:
\begin{align*}
\Prob_0(\bY_m \leq r , \tau_{r^*} \leq m) \leq (r+1) m e^{-\frac{n\delta}{2}+1}.
\end{align*}
\end{lemma}
\begin{proof}
Denote the \textit{last} hitting time of $r^*$ before time $m$ by
\begin{align*}
    \tau^{l,m}_{r^*} := \sup\{m': \bY_{m'} = r^* \text{ and } \bY_{j} \neq r^* \text{ for all } j \text{ such that } m' < j \leq m  \}.
\end{align*}
By the law of total probability
\begin{align*}
     \Prob_0(\bY_m \leq r , \tau_{r^*} \leq m) &= \sum^{m-1}_{m'=1}\Prob_0(\bY_m \leq r , \tau^{l,m}_{r^*} = m') \\
     &= \sum^{m-1}_{m'=1}\Prob_0(\bY_m \leq r , \bY_{m-1} \neq r^*, \ldots, \bY_{m'+1} \neq r^* \mid \bY_{m'} = r^* ) \Prob_{0}(\bY_{m'} = r^*) \\
     &\leq (r+1) m e^{-\frac{n\delta}{2} + 1},
\end{align*}
where the inequality follows by Lemma~\ref{lemma:stationary bound BD process} since for any $m' \in \{1,\ldots,m-1\}$
%
\begin{multline*}
    \Prob_0(\bY_m \leq r , \bY_{m-1} \neq r^*, \ldots, \bY_{m'+1} \neq r^* \mid \bY_{m'} = r^* ) \\
    = 
    \sum^{r}_{r' = 0} \Prob_0(\bY_m = r' , \bY_{m-1} \neq r^*, \ldots, \bY_{m'+1} \neq r^* \mid \bY_{m'} = r^* ) \leq (r+1) e^{-\frac{n\delta}{2} + 1}.
\end{multline*}
\end{proof}

With Lemmas~\ref{lemma:first hit BD process}~and~\ref{lemma:return time BD process} in hand, we now use \eqref{Eqn:Yrdecomp} to bound \ref{eqn:Second} in \eqref{eqn:proof decomp}. Recall that $M_1(t) :=  \{\lfloor \frac{n\lambda t q}{2} \rfloor+1,\ldots,p(n) \}$, where $p(n)$ is a polynomial in $n$ that we will choose later.
\begin{lemma}
\label{lemma:second likelihood UB}
Let $\delta^{*}(t) = \delta^* := \min\left(\delta_1(t),\delta_2 \right)$, with $\delta_1(t)$ and $\delta_2$ defined in Lemmas~\ref{lemma:first hit BD process} and \ref{lemma:stationary bound BD process}, respectively. Let 
\begin{align*}
    r^{**} = \lfloor n\delta^{*}(t)  \rfloor,
\end{align*}
and assume $r < r^{**} / 2 =  \lfloor n \delta^*(t) \rfloor/2$.
%
%
There exists a time $t_0 \in (0,1)$ and constants $c_0,c_1 \in (0,\infty)$ that depend only on $\gmax, \gmin, q, \lambda$ such that if $t \in (0,t_0)$ 
%
%
and $T \in (t,\infty)$, then 
%
\begin{align*}
e^{-Tn q \lambda}\sum_{m \in M_1(t) } \frac{(n T q \lambda)^m}{m!} \tR_{\x,\y}^m  \leq  e^{-tn c_0 + c_1\log(n)}.
\end{align*}
\end{lemma}
\begin{proof}
Recall that by $\tR^m_{\x,\y} \leq \Prob_0(\bY_m \leq r)$ by Lemma~\ref{lemma:BD process minor}. Since $r < \lfloor  \delta^*(t) \rfloor / 2$ and $\delta^*(t)  = \min(\delta_1(t),\delta_2)$, both  Lemmas~\ref{lemma:first hit BD process} and \ref{lemma:return time BD process} apply and we obtain using \eqref{Eqn:Yrdecomp}
\begin{align*}
\Prob_0(\bY_m \leq r) \leq \Prob_0(\tau_{r^{**}} > m) + \Prob_0(\bY_m \leq r,\tau_{r^{**}} \leq m) \leq e^{-\frac{m \tgmin^2}{8\lambda^2}} +  (r+1) m e^{-\frac{n\delta^{*}}{2}+1}.
\end{align*}
It follows that
\begin{align*}
e^{-Tn q \lambda}\sum_{m \in M_1(t) } \frac{(n T q \lambda)^m}{m!}\tR_{\x,\y}^m  
& \leq e^{-Tn q \lambda}\sum_{m \in M_1(t) } \frac{ ( n T q \lambda)^m}{m!} \Big[ \big(e^{\frac{-\tgmin^2}{8 \lambda^2}} \big)^m  + (r+1)m e^{-\frac{n\delta^{*}}{2}+1}\Big]\\
&\leq e^{-Tn q \lambda} e^{-nTq\lambda e^{-\frac{\tgmin^2}{8\lambda^2} } } + e^{-Tn q \lambda} (r+1) p(n) e^{-\frac{n\delta^{*}}{2}+1} \\
&\leq e^{-tnq\lambda(1-e^{-\frac{\tgmin^2}{8\lambda^2}} })   + (r+1) p(n)e^{-\frac{n\delta^{*}}{2}+1},
\end{align*}
%
where the second inequality used that $m < p(n)$ for all $m \in M_1(t)$, and the third inequality used that $T > t$. Now let
\begin{align}
\label{eqn:t star def}
t_0 =  \min\left( \frac{1}{\tgmax + q\tgmin}, \frac{q}{q + e \tg_* }  \right).
\end{align}
If $t < t_0$ then 
\begin{align}\label{eqn:delta star LB}
\delta^{*}(t) = \min\left(\frac{tq\tgmin}{2t (\tgmax + q\tgmin}) + 4, \frac{q}{q + e\tg_* } \right) &> t\min\left(\frac{q\tgmin}{6}, \frac{q t^{-1}}{q + e\tg_* } \right) \\
&> t \min\left(\frac{q\tgmin}{6},1\right),
\end{align}
where the first inequality follows since $t < t_0 < (\tgmax + q\tgmin)^{-1}$ and the second since $t^{-1} > (t_0)^{-1} > (q / (q+e\tg_*))^{-1}$. Therefore, we obtain for $t < t_0$
\begin{align*}
    e^{-Tn q \lambda}\sum_{m \in M_1(t) } \frac{(n T q \lambda)^m}{m!}\tR_{\x,\y}^m &\leq e^{-tnq\lambda(1-e^{-\frac{\tgmin^2}{8\lambda^2} }) }  + (r+1) p(n)e^{-\frac{n\delta^{*}}{2}+1} \\
    &\leq e^{-tnq\lambda(1-e^{-\frac{\tgmin^2}{8\lambda^2} }) }  + (r+1) p(n)e^{-tn \min\left(\frac{q\tgmin}{12},\frac{1}{2}\right)+1}.
\end{align*}
Hence the stated bound follows for sufficiently small $c_0 \in (0,\infty)$ and sufficiently large $c_1 \in (0,\infty)$.
\end{proof}
\subsection{Bounding \ref{eqn:Third} Using Stationarity of $\bY$}
We now bound \ref{eqn:Third} in \eqref{eqn:proof decomp}. Again recall that it suffices to bound $\Prob_0(\bY_m \leq r)$ since by \eqref{eqn: Hamming distance relationship} and Lemma~\ref{lemma:BD process minor} 
\begin{align}
\label{eqn:spec gap overview}
\R^m_{\x,\y} \leq \Prob_0(\bX_m = r) \leq \Prob_0(\bX_m \leq r) \leq \Prob_0(\bY_m \leq r).
\end{align}
We will show that by suitable choice of $p(n)$ in \eqref{eqn:M sets 2}, we have $\mathcal{L}(\bY_m) \approx \nu$ for $m \in M_2$, that is, $\{\bY_m\}$ is close to stationarity. Hence, a bound on $\Prob(\bY_m \leq r)$ for $m \in M_2$ can be obtained by bounding $\sum^r_{j=0}\nu(j)$.  
We will use the following result (see equation (12.13) in \cite{Levin:2009})
Since $\bB$ is $\nu$-reversible  
\begin{align*}
 \bB^m_{0,k} \leq \nu(k) + \frac{e^{-\text{SpecGap}(\bB) m }}{\min_x\nu(x)},
\end{align*}
where $\text{SpecGap}(\bB) = 1 - \lambda_2$ for $\lambda_2$ the second largest eigenvalue of $\bB$. This implies
\begin{align}
\label{eqn:spec gap bound}
\Prob_0(\bY_m \leq r) = \sum^r_{k=0} \bB^m_{0,k} \leq \sum^r_{k=0}\nu(k) + \frac{re^{-\text{SpecGap}(\bB) m }}{\min_x\nu(x)} .
\end{align}
%


Lemma~\ref{lemma:BD stationary} allows us to bound $\sum^r_{k=0}\nu(k)$ and $\min_x\nu(x)$. 
It remains to bound the $e^{-\text{SpecGap}(\bB) m}$ term in \eqref{eqn:spec gap bound}. To do this, we will obtain a lower bound on $\text{SpecGap}(\bB)$ using the canonical path method \cite{Diaconis:1991}. Note that tighter, but less explicit, bounds for birth-death chains are given in \cite{Chen:2013}. 
\begin{theorem}(
Proposition 1'
\citet{Diaconis:1991})
\label{thm:canonical paths thm}
Let $\bB$ be the transition matrix of a Markov chain defined on a finite state space $\X$ with stationary distribution $\nu$. 
Let $\gamma_{x,y}$ denote a directed path between each element $x,y \in \X$ with $x \neq y$,where a path is a sequence edges (pairs $(a,b)$ for $a,b \in \mathcal{X}$ with $\nu(a)\bB_{a,b} > 0$) such that each edge appears at most 
once, but vertices may repeat.
Then
\begin{align}
\text{SpecGap}(\bB) \geq \left(\max_{a \neq b: \bB_{a,b} \neq 0} \frac{1}{\nu(a) \bB_{a,b}} \sum_{x \neq y: (a,b) \in \gamma_{x,y}} \nu(x) \nu(y) |\gamma_{x,y}| \right)^{-1},
\label{Eqn:GapBbound}
\end{align}
where $|\gamma_{x,y}|$ denotes the number of edges along the path $\gamma_{x,y}$.
\end{theorem}

We will use Theorem~\ref{thm:canonical paths thm} to prove the following result.
\begin{lemma}
\label{lemma:BD Spectral}
The following bound holds:
\begin{align*}
\text{SpecGap}(\bB) \geq \frac{\min(\tgmax q^{-1},\tgmin )}{\lambda n^{3}(n+1)}.
\end{align*}
\end{lemma}
\begin{proof}
We bound each term in \eqref{Eqn:GapBbound} in turn.

Let $x,y \in \{0,\ldots,n\}$ and 
$\gamma_{x,y} = ((x,x+1), (x+1,x+2),\ldots,(y-1,y))$ for $x < y$, and $\gamma_{x,y}$ defined similarly for $y > x$.
In this case $|\gamma_{x,y}| \leq n$.

A uniform lower bound on the transition probability $\bB_{a,b}$ is obtained at the transitions to the $\{0,n\}$ states (see \eqref{eqn:transition probs}):
\begin{align}
\label{eqn:Gap chain proof 1}
\bB_{i,j} \geq \frac{1}{n\lambda} \min\left(\frac{\tgmax}{q},\tgmin \right) \text{ for all } i \neq j \text{ such that }|i - j| \leq 1.
\end{align}
Let $a,b \in \{0,\ldots,n\}$ and consider any path $\gamma_{x,y}$ such that the edge $(a,b) \in \gamma_{x,y}$. Since $\gamma_{x,y}$ is the simple path from $x$ to $y$, any vertex visited along $\gamma_{x,y}$ must lie between $x$ and $y$ and so
\begin{align*}
x \leq a \leq y \quad \text{  if  } x<y, \quad\quad y \leq a \leq x \quad \text{  if  } y < x .
\end{align*}
To bound $\nu(x)\nu(y)/\nu(a)$,  
without loss of generality suppose $x < y$ and so $x \leq a \leq y$. Recall that $\nu$ is a binomial distribution with mode $\lfloor (n+1) \frac{q}{\tg_*+q} \rfloor$. Hence,
\begin{align}
\label{eqn:Spec Gap Proof Ordering}
\nu(a) \geq \nu(x) \text{ if } a \leq \lfloor (n+1) \frac{q}{\tg_*+q} \rfloor \quad\quad \nu(a) \geq \nu(y) \text{ if } a \geq \lfloor (n+1) \frac{q}{\tg_*+q} \rfloor, 
\end{align}
which implies that for any vertex $a$ that lies inside $\gamma_{x,y}$ we have $\nu(x)\nu(y)/\nu(a) \leq 1$. 

Finally, note that there are in total $n(n+1)$ paths since there are $|\{0,\ldots,n\} | = n+1$ states, with a directed path to each of the other $n$ states. Hence,
\begin{align}
\label{eqn:Gap chain proof 2}
\sum_{x \neq y: (a,b) \in \gamma_{x,y}} \frac{\nu(x) \nu(y)}{\nu(a)} \leq    n(n+1).
\end{align}
Since $a$ and $b$ were arbitrary, we finally have by \eqref{eqn:Gap chain proof 1}, \eqref{eqn:Gap chain proof 2}, and using $|\gamma_{x,y}| \leq n$ that
\begin{align}
\label{eqn:Gap chain proof end}
   \max_{a \neq b : \bB_{a,b} \neq 0} \frac{1}{\nu(a) \bB_{a,b}} \sum_{x \neq y: (a,b) \in \gamma_{x,y}} \nu(x) \nu(y) |\gamma_{x,y}| \leq \frac{\lambda n^{3}(n+1)}{\min(\tgmax q^{-1},\tgmin )}.
\end{align}
The stated bound follows by Theorem~\ref{thm:canonical paths thm} using \eqref{eqn:Gap chain proof end}.
\end{proof}
With the lower bound on $\text{SpecGap}(\bB)$ in hand, we now use \eqref{eqn:spec gap bound} to establish an upper bound on  \ref{eqn:Third} in \eqref{eqn:proof decomp} using the following lemma. Recall that the function $p(n)$ introduced in \eqref{eqn:M sets 2} defines the lower bound $m \geq p(n)$ for all $m \in M_2$. Our choice of $p(n)$ will be the upper bound on the relaxation time of $\bB$ obtained from Lemma~\ref{lemma:BD Spectral}. 
%
\begin{lemma}
\label{lemma:third likelihood UB}
Assume $\frac{r}{n} < \frac{q}{ 2(q+\tg_*)}$. There exists a constant $c_0 \in (0,\infty)$ that depends only on $\tgmin$, $\tgmax$, $\lambda$, $q$, such that by setting $p(n) = c_0 n^5 = \bigO(n^5)$
%
%
we have
\begin{align*}
e^{-n\lambda T q}\sum_{m \in M_2 } \frac{(n T q \lambda)^m}{m!}\tR_{\x,\y}^m \leq 2e^{-\frac{n}{2}(\frac{q}{\tg_*+q})^2}.
\end{align*}
\end{lemma}
\begin{proof}
Applying \eqref{eqn:spec gap overview} and \eqref{eqn:spec gap bound}
\begin{align*}
e^{-n\lambda T q}\sum_{m \in M_2 } \frac{(n T q \lambda)^m}{m!}\tR_{\x,\y}^m &\leq e^{-n\lambda T q}\sum_{m \in M_2 } \frac{(n T q \lambda)^m}{m!}\sum^r_{k=0} \bB^m_{0,k} \\
&\leq 
\sum^r_{k=0}\nu(k) + \frac{re^{-\text{SpecGap}(\bB) p(n) }}{\min_x\nu(x)}.
\end{align*}
%
where the final inequality follows since $m \geq p(n)$ for $m \in M_2$. Let \begin{align}\label{eqn:third likelihood pn}
 p(n) =\frac{2\lambda n^{5}\left( 2 +\log\left(\frac{\tg_*+q}{\min(\tg_*,q)}\right)  \right)}{\min\left(\tgmax q^{-1},\tgmin \right)} = \bigO(n^5).
\end{align}
By Lemmas~\ref{lemma:BD stationary} and \ref{lemma:BD Spectral} 
\begin{align}\label{eqn: proof bd spectral}
\frac{re^{-\text{SpecGap}(\bB) p(n) }}{\min_x\nu(x)} \leq r\left(\frac{(\tg_*+q)}{\min(\tg_*,q)}\right)^{n} \exp\left(-p(n) \frac{\min\left(\tgmax q^{-1},\tgmin \right)}{\lambda n^{3}(n+1)}\right) \leq e^{-n}.
\end{align}
To see this, note that to establish the second inequality in \eqref{eqn: proof bd spectral} it suffices to choose $p(n)$ such that
\begin{align}\label{eqn: proof bd spectral 1}
    \left(n + \log(r) + n\log\left(\frac{\tg_*+q}{\min(\tg_*,q)}\right)\right) \lambda n^{3}(n+1) \leq p(n) \min\left(\tgmax q^{-1},\tgmin \right),
\end{align}
and using $n^{3}(n+1) \leq 2n^{4}$, $n + \log(r) \leq 2n$, we have that the left-hand side in \eqref{eqn: proof bd spectral 1} is bounded above as
\begin{align*}
\left(n + \log(r) + n\log\left(\frac{\tg_*+q}{\min(\tg_*,q)}\right)\right) \lambda n^{3}(n+1) \leq 2\lambda n^{5}\left( 2 +\log\left(\frac{\tg_*+q}{\min(\tg_*,q)}\right)  \right).
\end{align*}
Now applying Lemma~\ref{lemma:BD stationary} to bound $\sum^r_{k=0}\nu(k)$ we obtain: 
\begin{align*}
e^{-n\lambda T q}\sum_{m \in M_2 } \frac{\tR_{\x,\y}^m (n T q \lambda)^m}{m!} \leq \sum^r_{k=0}\nu(k) + \frac{re^{-\text{SpecGap}(\bB) p(n) }}{\min_x\nu(x)} &\leq  e^{-\frac{n}{2}\left(\frac{q}{\tg_*+q}\right)^2} +  e^{-n}  \\
&\leq  2e^{-\frac{n}{2}(\frac{q}{\tg_*+q})^2}.
\end{align*}
\end{proof}

\subsection{Proof of Theorem~\ref{thm:main thm appdx A}}
The following lemma is the final piece to prove \eqref{eqn:overview statement} and provides an upper bound on the numerator in \eqref{eqn:TRatioBound} required to prove  Theorem~\ref{thm:main thm appdx A}.
\begin{lemma}
\label{lemma:main lemma appdx a}
Let $t \in (0, t_0)$, with $t_0$ the constant from \eqref{eqn:t star def}.
Assume $\lambda \geq \max(\tgmax,\tgmin + q^{-1}\tgmax)$ and $ r < \lfloor n\delta^{*}(t)\rfloor/2$, with
$\delta^{*}(t)$ 
%
%
defined as in Lemma~\ref{lemma:second likelihood UB}.
%
%
There exists constants $c_0,c_1,c_2 \in (0,\infty)$ that depend only on $\tgmin,\tgmax,\lambda$ and $q$ such that 
%
%
%
\begin{align*}
\sup_{T \in (t,\infty)}p_{(T,\tQ)}(\y \mid \x) \leq e^{-nt c_0 + \log(n) c_1+ c_2}.
\end{align*}
\end{lemma}

\begin{proof}
Recall the expansion in \eqref{eqn:proof decomp}:
\begin{align}
p_{(T,\tQ)}(\y \mid \x) &= e^{-Tn q \lambda}\left[\sum_{m \in M_0(t) } \frac{\tR_{\x,\y}^m (n T q \lambda)^m}{m!} + \sum_{m \in M_1(t) } \frac{\tR_{\x,\y}^m (n T q \lambda)^m}{m!}\right]  \\
&+ e^{-Tn q \lambda}\sum_{m \in M_2 } \frac{\tR_{\x,\y}^m (n T q \lambda)^m}{m!} \label{eqn: proof bound ii} .
\end{align}
Assume $T \in (t,\infty)$ and we choose $p(n)$ as in Lemma~\ref{lemma:third likelihood UB} so that the bound on \eqref{eqn: proof bound ii} given in Lemma~\ref{lemma:third likelihood UB} holds; recall that $p(n) = \bigO(n^5)$ in this case. To apply the bounds given in Lemmas~\ref{lemma:second likelihood UB} and Lemma~\ref{lemma:third likelihood UB}, we require that $r < \lfloor n\delta^{*}(t)\rfloor/2$ and $ r < \frac{nq}{2(q + \tg_*)}$, respectively. The former is true by assumption and it is straightforward to check that $ r < \lfloor n\delta^{*}(t)\rfloor/2$ also implies $r < \frac{nq}{2(q + \tg_*)}$ as well. Now applying Lemma~\ref{lemma:first likelihood UB}, Lemma~\ref{lemma:second likelihood UB}, and Lemma~\ref{lemma:third likelihood UB}, and recalling that $t < t_0$ by assumption, we have that there exists model-dependent constants $c,c' \in (0,\infty)$ such that
\begin{align*}
p_{(T,\tQ)}(\y \mid \x) \leq e^{-\frac{nt q \lambda}{8}} +  e^{-tnc + \log(n) c^{\prime}} + 2e^{-\frac{n}{2}(\frac{q}{\tg_*+q})^2}.
\end{align*}
Hence, there exists sufficiently small $c_0 \in (0,\infty)$ and sufficiently large $c_1,c_2 \in (0,\infty)$ such that $p_{(T,\tQ)}(\y \mid \x) \leq e^{-nt c_0 + \log(n) c_1 + c_2}$ provided $t < t_0$. Since this inequality holds for any $T \in (t,\infty)$, we obtain the stated bound.
%
%

\end{proof}
We are now in a position to complete the proof of Theorem~\ref{thm:main thm appdx A}.
\begin{proof}(Proof of Theorem~\ref{thm:main thm appdx A})



Recall $J := [a\frac{r}{n},b\frac{r}{n}]$ for constants $0 < a < b$ and suppose $T^* \in J$.
For the moment, assume $r < \lfloor n\delta^{*}(t) \rfloor/2$ so that Lemma~\ref{lemma:main lemma appdx a} applies with $t$ chosen to be
\begin{align*}
t = \min(t_0, h(n)) = \min\left( \frac{1}{\tgmax + q\tgmin},\frac{q}{q + e \tg_* }, h(n) \right),
\end{align*}
where $h(n)$ is a function that we will choose momentarily. Lemma~\ref{lemma:likelihood LB} provides us with the lower bound
\begin{align*}
p_{(T^{*},\tQ)}(\y \mid \x) &\geq \exp\left(-nT^{*} q\lambda + r \log(T^{*} \tgmin) \right) \\
&\geq \exp\left(-r q\lambda b  + r \log(a r \tgmin) - r\log(n) \right) \\
&= \exp\left(-r\log(n) (1 + q\lambda b)  + r \log(a r \tgmin)  \right),
\end{align*}
provided $n \geq 3$. Since this bound holds for any $T^* \in J$, we have
\begin{align}\label{eqn: main theorem proof inf bound}
    \inf_{T^* \in J} p_{(T^{*},\tQ)}(\y \mid \x) &\geq \exp\left(-r\log(n) (1 + q\lambda b)  + r \log(a r \tgmin)  \right).
\end{align}
 By Lemma~\ref{lemma:main lemma appdx a}, there exists model-dependent constants $c_0,c_1,c_2 \in (0,\infty)$ such that the following upper bound holds: 
%
\begin{align}\label{eqn: main theorem proof sup bound}
\sup_{T \in (t,\infty)}p_{(T,\tQ)}(\y \mid \x) &\leq \exp(-tn c_0 + \log(n)c_1 + c_2) \\
&= \exp(-\min(t_0,h(n))n c_0 + \log(n)c_1 + c_2).
\end{align}
Choosing $h(n) = \frac{2r\log(n)(1+q\lambda b)}{n c_0}$, we have that $t = \min(t_0,h(n)) = h(n)$ for 
\begin{align*}
    r \leq \frac{2t_0 c_0 n }{\log(n)(1+q\lambda b)},
\end{align*}
in which case by \eqref{eqn: main theorem proof sup bound} and our choice of $h(n)$,
\begin{align}\label{eqn: main theorem proof sup bound 2}
  \sup_{T \in (t,\infty)}p_{(T,\tQ)}(\y \mid \x) \leq  \exp(-2r \log(n)(1+q\lambda b) + \log(n) c_1 + c_2).
\end{align}
Combining the bounds \eqref{eqn: main theorem proof inf bound} and \eqref{eqn: main theorem proof sup bound 2}, we have that there exists a constant $c_3 > 0$ that depends on $\gmin,\gmax, \lambda, q, a, b$ such that for $r$ and $n$ sufficiently large
\begin{align}
\label{eqn: thm bound statement}
\frac{\sup_{T \in (t,\infty)}p_{(T,\tQ)}(\y \mid \x)}{\inf_{T^* \in J}p_{(T^{*},\tQ)}(\y \mid \x)} = \frac{\sup_{T \geq h(n)}p_{(T,\tQ)}(\y \mid \x)}{\inf_{T^* \in J}p_{(T^{*},\tQ)}(\y \mid \x)} \leq e^{-c_3 r \log(n) }.
\end{align}
The bound stated in Theorem~\ref{thm:main thm appdx A} follows by \eqref{eqn: thm bound statement}. To see this, note  
\begin{align*}
    h(n) =\frac{2r\log(n)(1+q\lambda b)}{n c_0} = c_4  \frac{r}{n} \log(n)  \quad \text{ with } \quad c_4 = \frac{2(1+q\lambda b)}{c_0}
\end{align*}
and so for $I = (0, c_4 \frac{r}{n}  \log(n))$, we have by \eqref{eqn: thm bound statement} 
\begin{align*}
    \frac{\sup_{T  \in I}p_{(T,\tQ)}(\y \mid \x)}{\inf_{T^* \in J}p_{(T^{*},\tQ)}(\y \mid \x)} \leq e^{-c_3 r \log(n) }.
\end{align*}
Finally, it remains to verify that $r < \lfloor n\delta^{*}(t) \rfloor/2$. By our choice of $h(n)$, we recall $t = \min(t_0, h(n)) = h(n)$ and so 
\begin{align*}
\lfloor n\delta^{*}(t) \rfloor/2 = \lfloor n\delta^{*}(h(n)) \rfloor/2 = \bigO(r\log(n)).
\end{align*}
Hence, $r < \lfloor n\delta^{*}(t) \rfloor/2$ holds for $n$ sufficiently large, completing the proof.
%
%

\end{proof}

\subsection{Proof of Theorem~\ref{thm: T bound symmetric}}
We can improve the lower bound on $T$ given in Theorem~\ref{thm:main thm appdx A} from $\bigO(r\log(n)/n)$ to $\bigO(r/n)$ under further model assumptions. Recall
\begin{align}
\tR^m_{\x,\y} = \Prob(X_m = \y) = \Prob(X_m = \y \mid \bX_m = r)\Prob(\bX_m = r ),
\end{align}
which implies the inequality $\tR^m_{\x,\y} \leq \Prob(\bX_m = r )$ as in \eqref{eqn: Hamming distance relationship}. Thus our analysis ignored the term $\Prob(X_m = \y \mid \bX_m = r)$, which leads to an additional $\log(n)$ term. The proof of Theorem~\ref{thm: T bound symmetric} follows by establishing a modified version of Theorem~\ref{thm:main thm appdx A} for symmetric evolution models. 

\begin{proof}(Proof of Theorem~\ref{thm: T bound symmetric}) \\

Consider a symmetric evolution model where sites have equal rates $\tgmax = \tgmin = \gamma \in (0,\infty)$. Let $J = [a\frac{r}{n}, b\frac{r}{n}]$ for constants $0 < a < b$. Using Lemma~\ref{lemma:likelihood LB} and that $a\frac{r}{n} \leq T^* \leq b\frac{r}{n}$ for $T^* \in J$, we have
\begin{align}\label{eqn: T bound symmetric lower bound}
    \inf_{T^* \in J} p_{(T^{*},\tQ)}(\y \mid \x) &\geq \exp\left(-nT^{*} q\lambda + r \log(T^{*} \gamma) \right) \\
    &\geq \exp\left(-br q\lambda + r\log(\frac{rq}{n}) + r\log(\frac{a \gamma}{q}) \right).
\end{align}
%
We will now obtain an upper bound on $p_{(T,\tQ)}(\y \mid \x)$ that is an improvement over the one in \eqref{eqn: main theorem proof sup bound}. Define the Hamming ball of radius $r$ centered at $\x$ by $\mathcal{H} = \{\z: \text{d}_{\text{H}}(\x,\z) = r \}$. Then
\begin{align}\label{eqn: T bound symmetric proof improvement}
\Prob(X_m = \y \mid \bX_m = r) = \frac{\tR^m_{\x,\y}}{\sum_{\mathcal{H}} \tR^m_{\x,\z}} = \frac{1}{|\mathcal{H}|} = \frac{1}{\binom{n}{r}q^r} \leq e^{r\log(\frac{rq}{n})},
\end{align}
where we used that $\tR^m_{\x,\y} = \tR^m_{\x,\z}$ for $z \in \mathcal{H}$ by symmetry and $\binom{n}{r}q^r \geq (\frac{nq}{r} )^r$. The argument for the improved upper bound on $p_{(T,\tQ)}(\y \mid \x)$
is the same, but now includes the previously unaccounted for $\Prob(X_m = \y \mid \bX_m = r)$ term. Specifically, assume as in the proof of Theorem~\ref{thm:main thm appdx A} that $r < \lfloor n\delta^{*}(t) \rfloor/2$ so that Lemma~\ref{lemma:main lemma appdx a} applies and set
\begin{align*}
t = \min(t_0, h(n)) ,
\end{align*}
with $t_0$ the constant from \eqref{eqn:t star def}. Here, we choose
\begin{align*}
    h(n) = \frac{r}{n}\max\bigg\{\frac{2}{c_0}(q\lambda b + \log(\frac{q}{a\gamma})), \frac{4}{\min(q\tgmin/6,1)} \bigg\},
\end{align*}
where $c_0 \in (0,\infty)$ is the constant in Lemma~\ref{lemma:main lemma appdx a}. Note that $t = h(n)$ for $r/n$ sufficiently small. Hence Lemma~\ref{lemma:main lemma appdx a} together with \eqref{eqn: T bound symmetric proof improvement} gives that for $n$ large enough
\begin{align}\label{eqn: T bound symmetric upper bound}
\sup_{T \in (t,\infty)}p_{(T,\Q)}(\y \mid \x) = \sup_{T \geq h(n)}p_{(T,\Q)}(\y \mid \x) \leq e^{-n h(n) c_0 + r\log(\frac{rq}{n})  + \log(n) c_1 + c_2},
\end{align}
for model-dependent constants $c_1,c_2 \in (0,\infty)$. Combining \eqref{eqn: T bound symmetric lower bound} and \eqref{eqn: T bound symmetric upper bound} and using $h(n) \geq \frac{r}{n}\frac{2}{c_0}(q\lambda b + \log(\frac{q}{a\gamma}))$, it follows that there exists a model-dependent constant $c_3 > 0$ that depends on $\gamma, \lambda, q, a, b$ such that for $n$ large enough
\begin{align}\label{eqn: T bound symmetric main statement}
    \frac{\sup_{T \geq h(n)}p_{(T,\Q)}(\y \mid \x)}{\inf_{T \in J}p_{(T,\Q)}(\y \mid \x)} \leq e^{-c_3r},
\end{align}
provided $r/\log(n) \rightarrow \infty$. Theorem~\ref{thm: T bound symmetric} follows by \eqref{eqn: T bound symmetric main statement} and choosing $ I = (0,h(n)) = (0, \frac{r}{n}c_4)$, with
\begin{align*}
     c_4 = \max\bigg\{\frac{2}{c_0}(q\lambda b + \log(\frac{q}{a\gamma})), \frac{4}{\min(q\tgmin/6,1)} \bigg\}.
\end{align*}
To complete the proof, we need to check that $r < \lfloor n\delta^{*}(t) \rfloor/2$.  For $n$ large enough, we have $t = h(n) < t_0$, in which case by \eqref{eqn:delta star LB} we have $\delta^{*}(\jrm{t}) \geq t\min(\frac{q\gamma_{\min}}{6},1)$. using $\lfloor x \rfloor \geq x - 1$ we obtain
\begin{align}\label{eqn: T bound symmetric verify}
\lfloor n \delta^{*}(t) \rfloor/2 \geq n\delta^{*}(t)/2 - 1/2 \geq \frac{nt}{2}\min\left(\frac{q\tgmin}{6},1 \right) - \frac{1}{2}.
\end{align}
Hence, using $t = h(n)$ for $n$ sufficiently large and that $h(n) \geq \frac{r}{n} \frac{4}{\min(q\tgmin/6,1)}$, we have by \eqref{eqn: T bound symmetric verify} that for $n$ large enough
\begin{align*}
\lfloor n \delta^{*}(t) \rfloor/2 \geq \frac{nt}{2}\min\left(\frac{q\tgmin}{6},1 \right) - \frac{1}{2} &= \frac{n h(n)}{2} \min\left(\frac{q\tgmin}{6},1 \right) - \frac{1}{2n} \\
 &\geq 2r - \frac{1}{2n} > r.
\end{align*}
It follows that $r < \lfloor n \delta^{*}(t) \rfloor/2 $, completing the proof.
\end{proof}

\section*{Data Availability}
Not applicable.

\bibliography{Bibliography}

\appendix

\section{Proof of Lemma~\ref{lemma:return time bound}}
We first restate the lemma here for readability.
\hittingtime* 
\noindent The proof of Lemma~\ref{lemma:return time bound} uses the following result given in \citet{Levin:2009}.
\begin{proposition}\label{prop: levin and peres}(\cite{Levin:2009} Proposition 1.14) Consider the Markov chain $\{Z_t\}$ in Lemma~\ref{lemma:return time bound} with stationary distribution $\nu$. 
For $i,j \in \X$, let $N_{i,j}$ be the number of visits to state $j$ before returning to $i$:
\begin{align*}
    N_{i,j} := \sum^{\infty}_{t=0} \mathbbm{1}(Z_t = j, \tau^{+}_{i} > t).
\end{align*}
Then $ \nu(j) = \frac{\E_i[N_{i,j}]}{\E_i[\tau_i^+]}$ provided $\E_i[\tau_i^+] < \infty$, where $\E_i$ denotes the expectation when the chain is initialized at state $i$.
\end{proposition}
\noindent The proof of Lemma~\ref{lemma:return time bound} follows by Proposition~\ref{prop: levin and peres} and Markov's inequality.
\begin{proof}(Proof of Lemma~\ref{lemma:return time bound})
Applying Proposition~\ref{prop: levin and peres} with $i = j$ gives us the identity $\nu(i) \E_i[\tau^+_i] = 1$ since $\E_i[N_{i,i}] = 1$. Applying this identity gives us for any $i,j \in \X$
\begin{align*}
     \E_i[N_{i,j}] = \E_{i}[\tau_i^+]\nu(j) = \frac{\E_{i}[\tau_i^+]\nu(j)}{\E_{i}[\tau_i^+]\nu(i)} = \frac{\nu(j)}{\nu(i)}.
\end{align*}
Now let $r,r^* \in \X$. Using $\{\tau_r < \tau_{r^*}^+ \} = \{ N_{r^{*},r} \geq 1\}$ and Markov's inequality gives us the stated bound:
\begin{align*}
     \Prob(\tau_{r} < \tau^{+}_{r^*} \mid Z_0 = r^*) =  \Prob(N_{r^*,r} \geq 1 \mid Z_0 = r^*) \leq \E_{r^*}[N_{r^{*},r}] = \frac{\nu(r)}{\nu(r^*)}.
\end{align*}
\end{proof}

\end{document}